\documentclass[fleqn]{mnras}
\usepackage{graphicx}
\usepackage[fleqn]{amsmath}
\usepackage{amssymb}
\usepackage{bm}
\usepackage{hyperref}
\hypersetup{
    colorlinks=true,                          
    linkcolor=black, % Couleur des liens internes
    citecolor=black, % Couleur des numéros de la biblio dans le corps
    urlcolor=blue} % Couleur des url
\title[A pair of giant planets in mean motion resonance]{On the orbital evolution of a pair of giant planets in mean motion resonance}

\author[Q. Andr\'e and J.C.B. Papaloizou]{Q. Andr\'e$^{1,2}$\thanks{E-mail: quentin.andre@ens-cachan.fr} and J.C.B. Papaloizou$^{1}$\\
$^{1}$ DAMTP, University of Cambridge, Wilberforce Road, Cambridge CB3 OWA, United Kingdom\\
$^{2}$ D\'epartement de Physique, ENS Cachan, Universit\'e Paris-Saclay, 61 Avenue du Pr\'esident Wilson, 94230 Cachan, France
}

\date{Accepted 2016 June 30. Received 2016 June 30; in original form 2015 July 09}

\pubyear{2016}

\begin{document}
\label{firstpage}
\pagerange{\pageref{firstpage}--\pageref{lastpage}}
\maketitle

\begin{abstract}
{Pairs of  extrasolar giant planets in a mean motion commensurability are common 
with 2:1 resonance occurring most frequently. Disc-planet interaction 
provides a mechanism for their origin. However, the time scale on which this could  operate
in particular cases is unclear.
We  perform  2D and 3D  numerical simulations of pairs of giant planets 
in a protoplanetary disc as they form and maintain  a mean motion commensurability.
We consider systems with current parameters similar to those of HD 155358, 24 Sextantis and HD 60532, 
and disc models of varying mass, decreasing mass corresponding to increasing age. %54.
For the lowest mass discs, systems with planets in the Jovian mass range 
migrate inwards  maintaining a 2:1 commensurability.
Systems with the inner planet currently at around 1 au from the central star 
could  have originated at a few  au  and migrated  inwards 
on a time scale comparable to protoplanetary disc lifetimes.
Systems of larger mass planets such as HD 60532 attain 3:1 resonance as observed.
For a given mass accretion rate, results are insensitive to the disc model for the range of viscosity prescriptions adopted, there being  good agreement between 2D and 3D simulations. %88
However, in a higher mass disc a pair of Jovian mass planets passes through 2:1 resonance
before attaining a temporary phase lasting a few thousand orbits in 
an unstable 5:3 resonance prior to undergoing a scattering.
Thus finding systems in this commensurability is unlikely.}
\end{abstract}

\begin{keywords} 
  accretion, accretion discs, hydrodynamics, methods: numerical, planetary systems,   planet-disc interactions, protoplanetary discs
\end{keywords}

\section{Introduction} \label{intro}
Pairs of  extrasolar giant planets in a mean motion commensurability are a common occurrence.
It has been estimated that sixth of  multi-planet systems  detected by the radial velocity technique
 are in  or close to a 2:1 commensurability
(Wright et al. 2011).
%Publications  of the Astronomical Society of the Pacific 123 (902): 412�. 
%Steffen et al.  The Astrophysical Journal Supplement Series 197 (1): 1�. arXiv:1102.0543. 
 Parameters for some cases  of interest  that are considered in this paper  are shown in table \ref{table1}
(for additional  examples  see
 e.g. Emelyanenko 2012).
  In addition there are two  known systems in 3:2  resonance { (Correia et al. 2009, Rein et al. 2012, Robertson et al. 2012a)}
 and one in a 4:3 resonance { (Johnson et al. 2011, Rein et al. 2012)}
 consistent with  the  systems in 2:1 resonance being the most  commonly observed commensurability.
 
 The existence of these resonant systems 
 indicates that dissipative mechanisms  that result in changes to planet semi-
major axes  that  produce related changes to period ratios in planetary systems 
 have operated.  This is  because  the  probability of forming
resonant configurations in situ is  expected to be small (e.g. Beauge et al. 2012).

 Disc-planet interaction can produce the required evolution of the semi-major axes.
 This may result in convergent migration leading to the formation of a commensurability
 (see Baruteau et al. 2014 and references therein). Accordingly understanding the  observed configuration of such systems
  has the potential for either revealing how disc-planet interactions may have operated or for ruling them out. 
  
 Previous { numerical} studies of commensurabilities forming 
 and evolving as a result of disc-planet interactions have focused on systems such as GJ 876, HD 45364 and HD 6805
 { interacting with disc modelled  with  either constant kinematic viscosity,
or  with the $\alpha$-viscosity parameter of  Shakura \& Sunyaev (1973) taken to be constant}
(for a review see Baruteau et al. 2014 and references therein). 
In this paper we extend such studies, considering systems with { orbital} parameters
similar to those of  HD 155358 {  (Robertson et al. 2012b)},  24 Sextantis {(Johnson et al. 2011)}, 
 HD 60532 { (Laskar \& Correia 2009)} 
{ as well as HD 6805 (Tifonov et al. 2014) each of} which have the inner planet
with semi-major axis in the range $0.5-1.4$ au.
We perform  2D and 3D  numerical simulations of pairs of giant planets 
interacting with a protoplanetary disc that attain a mean motion commensurability for up to $2.5 \times 10^4$ orbital periods of the inner planet. 
We investigate whether such systems 
could have originated at larger radii beyond the ice line 
and then migrated inwards, the commensurability possibly being formed in the same neighbourhood.

\begin{table}
 \begin{tabular}{c c c c c c}
\hline\hline                      
System &$ M_* $&$M_1 $ & $M_2$  &$ a_1 $ &$a_2$\\
  HD 155358&$0.92$&$0.85\pm0.05$&$0.82\pm0.07 $&$0.64$&$1.02$\\
  %HD 90043&&$1.99$&$0.86$&$1.33$&$2.08$\\
 24 Sextantis&$1.54$& $1.99\pm 0.4$& $ 0.86\pm 0.4$  &$1.33$&$2.08$\\
% HD 37124& $0.83$&$0.65\pm 0.05$  &$0.7\pm0.06$ &$1.7$&$2.8$\\
% HD 73526&1.08&$2.9\pm 0.2$&$2.5\pm 0.3$&$0.66$&$1.05$\\
%HD 82943& $1.2$&$2.01$&$1.75$&$0.75$&$2.15$\\
 %HD 128311&
 %HD 16069&$1.1$&$0.52$&$1.67$&$0.92$&$1.5$\\
  HD 6805&$1.7$& $2.5\pm 0.2$&$3.3\pm0.2$&$1.27$&$1.93$\\%e=0.13,0.1\pm0.05,0.06
  % 24 Sextantis&$1.54$& $1.99\pm 0.4$& $ 0.86\pm 0.4$  &$1.33$&$2.08$\\ 
 % NN Ser&$0.54$ &$2.28\pm0.4 $ &$6.91\pm 0.4$&$3.4$&$5.4$\\
   HD 60532&1.44 &$3.15$&$7.46$&$0.77$&$1.58$\\
\hline
\end{tabular}
\caption{Properties of the HD 155358, 24 Sextantis,  HD 6805  and HD 60532 systems.
The first three either are or possibly in  2:1  resonance while the fourth is in a 3:1 resonance. 
The first column identifies the system,
the second column gives the mass of the central star in solar masses.
The third and fourth columns give the masses of the planets in Jupiter masses  and the fifth and sixth columns give their semi-major axes in au.}
\label{table1}
\end{table}

In order to study the role of the nature of the underlying disc model we consider 
models with an inner MRI active region producing a significant effective viscosity and an outer inactive region 
for which a significant effective viscosity may occur only in the upper layers of the disc
(Gammie 1996) as well as models with a uniform $\alpha$-viscosity prescription throughout.
We also consider models with different surface density scaling corresponding to
varying the total disc mass or equivalently the steady state accretion rate.
In this way the disc-planet interaction at different stages of the life of the protoplanetary disc
can be studied with lower mass discs corresponding to later stages (e.g. Calvet et al.  2004).

We  find that  when  low mass disc models  are considered, 
systems with planets  in the Jovian mass range   maintain a  2:1  commensurability
while undergoing  inward type II migration.  This is found to be at a rate such that formation
at  a few  au from the central star and migration to their current locations 
on a time scale comparable to the expected protoplanetary disc lifetime
is possible in principle.

We find that there is  a relative insensitivity  of results to the disc model employed  and find  good agreement between 2D and 3D simulations.
Planets containing larger masses such as the HD 60532 system  which is observed to be in 3:1 resonance 
{(Laskar \& Correia 2009)} are found to 
attain this  resonance  in low mass low viscosity discs.  
 
We find that for systems with planets in the Jovian mass range, increasing the disc mass results in the formation of an unstable 5:3 resonance.
%and then a stable 3:2
%resonance. In the latter case inward migration is about an order of magnitude faster than for the lower mass discs 
%containing systems with a 2:1 commensurability. 
This  instability results in the rapid destruction  of the commensurability implying that the occurrence of such systems should be less common.

The plan of this paper is as follows.  We give  the basic equations and coordinate system used 
in Section \ref{Beq}. In Section \ref{Numsims} we outline the numerical methods  and computational domains
adopted going on to describe aspects of the physical set up and  disc models used in Sections \ref{discmod} and \ref{layered}.
We then indicate how results might be scaled to different radii and  summarise important aspects of type II migration in Sections \ref{scaling} and \ref{typeiimig}. 
We go on to describe our numerical results in Section \ref{Numres}, {beginning 
with  a comparison with previous results
for two migrating planets presented in section \ref{CompPreviousResults}. Finally,  
we discuss our conclusions in Section \ref{Disc}.}

\section{Basic Equations}\label{Beq}
We adopt a spherical coordinate system $(r,\theta,\varphi)$ with associated unit vectors
 $({\hat {\bm{r}} },{\hat {\bm{\theta}} },
 {\hat {\bm{\varphi}}}$) and origin at the centre of mass of the central star.

The basic equations governing the disc express the conservation of mass and momentum  under the gravitational potential due to the central star and any planets and incorporate a kinematic viscosity $\nu.$

\begin{flalign}
\frac{\partial \rho}{\partial t} &= -\bm{\nabla} \bm{\cdot} (\rho \bm{v}) \,\label{contg}, \\
\rho \frac{D\bm{v}}{D t}    
&= -\rho \bm{\nabla} \Phi   
-\bm{\nabla} P +\nabla \bm{\cdot} \bm{T}\, . 
\label{disk_eq}
\end{flalign}
Here the convective derivative is defined through
\begin{equation}
	\frac{D }{D t} \equiv \frac{\partial }{\partial t}+ \bm{v}\cdot\nabla,
\end{equation}
 where $\rho$ is the 
density, $\bm{v}$ is the velocity,
$P$ is the pressure, $\bm{T}$ is the viscous stress tensor (see e.g. Mihalas \& Weibel Mihalas 1984), and $\Phi$ is the gravitational potential which has contributions
from the central star and any planets present. Disc self-gravity is neglected.  
The pressure is related to the gas density and
the isothermal speed of sound $c_s$ through 
$P=\rho c_s^2.$

%The components of the
%viscous stress tensor $\bm{T}$  
%in Cartesian coordinates are given by 
%\begin{equation}
%T_{ik}=\rho \nu \left( \frac{\partial v_i}{\partial x_k} +
%\frac{\partial v_k}{\partial x_i} -\frac{2}{3} \delta_{ik} \del \bcdot \bm{v}
%\right) \ .
%\end{equation}

\section{Numerical Simulations}\label{Numsims}
Simulations were performed using the finite volume fluid dynamics code \texttt{PLUTO}
(Mignone et al. 2007) which has been used successfully to simulate protoplanetary discs interacting with planets
(e.g. Mignone et al. 2012,  Uribe et al. 2013). The planet positions are advanced using  a fourth order Runge-Kutta
method which however  assumes
%for numerical convenience
 that the forces due to the disc do not change 
 as  the planet locations are advanced through a time step,  making the method one of lower order
(see e.g. Nelson et al. 2000 for comparison). For some runs we employed the \texttt{FARGO} algorithm 
of Masset (2000) as this allows the numerical calculations to run significantly faster.
{We note that for our simulations  the options were chosen such that  algorithm was applied using the residual azimuthal velocity with respect to the initial azimuthal velocity, 
the latter  not being updated.
A sample of runs were checked carefully  to confirm numerical stability and that consistent results were obtained} (see also Mignone et al. 2012
for a comparison of this type).

For the calculations reported here, we adopt a locally isothermal equation of state for which
$c_s\propto r^{-1/2}.$ The constant of proportionality is chosen so as to give a constant aspect ratio
$h \equiv H/r=c_s/(r\Omega_K) = 0.05.$ Here $H$ is the disc semi-thickness and $\Omega_K$ is the local
Keplerian circular orbit angular velocity. We adopt a system of dimensionless units
such that masses are expressed in units of the central stellar mass $M_*,$ radii are expressed in units
of the initial orbital radius of the innermost planet and times are expressed in units of the
orbital period of a circular orbit at that radius. 
  
For three dimensional simulations the  radial computational  domain  in most cases was given  in dimensionless
units by  $r \in  [0.15,3.75].$ The grid outer boundary is treated as a rigid boundary and taken sufficiently far from the planets so that the density perturbations they create in the disc are damped before they reach it, while the grid inner boundary only allows inflow, so that the disc material can be accreted on to the central star. The $\theta$  domain was taken to be,  $[\pi/2-3H/r, \pi/2] \equiv [\theta_{\mathrm{min}}, \pi/2] ,$
with symmetry being assumed with respect to the plane $\theta =\pi/2,$ this being treated as a rigid  boundary.
 The $\varphi$  domain was taken to be  $[0, 2\pi].$  

%\onecolumn

%\twocolumn

The standard grid resolution for most simulations was taken to be  $(N_r,N_{\theta}, N_{\varphi}) = (162, 18, 314).$
The radial grid spacing was non uniform and chosen so that the grid spacing  $\Delta r$ was   equal to 
$0.02\,r.$ This geometric spacing is the most natural one, since the disc semi-thickness scales as $r$. The azimuthal grid spacing was uniform and such that $\Delta\varphi = \Delta r/r = 0.02.$
We remark that disc-planet interactions adopting  similar resolutions to those adopted here 
have been carried out for lower mass planets by Kley et al. (2009). The interval $\Delta \theta$ was chosen such that there were six grid cells per scale height.
Planet  gravitational potentials were softened by adopting a  gravitational softening
length taken to be $0.5$ Hill radii, being slightly less than two radial grid cells in extent,
and the smoothing filter of Crida et al. (2009a) was employed  when  calculating
torques acting on planets. We also performed { convergence}  checks using simulations  for which $N_r$ was increased to $229,$  $N_{\theta}$ increased to $25$ and $N_{\varphi}$ was increased to $444$, with the softening length reduced by a factor $\sqrt{2}.$

For two dimensional simulations,  in most cases the radial and azimuthal grids and domains
are the same as in the three dimensional case but now  $\theta$ is fixed to be $\pi/2.$
In this case the mass density $\rho$ is replaced by the surface density  $\Sigma$  and the pressure
is replaced  by a vertically integrated pressure. The gravitational softening
length was taken to be $0.6H$ and the smoothing filter of Crida et al. (2009a) was employed  when  calculating
torques acting on planets for this case also. The { convergence of two dimensional simulations was} checked by performing them at twice the resolution with the softening procedure remaining fixed.

As we consider  either low viscosity discs  or discs undergoing steady state accretion at very low 
accretion rates, we neglect accretion onto the planet. Many of our runs are carried out with planets 
migrating in discs with local kinematic viscosity $\nu < 10^{-6}$ in dimensionless units.
Accretion onto the planet has been found to be a small effect in this case (Bryden et al. 1999, Kley 1999).
In addition the time required for the accretion rate  through the disc to double the planet's mass 
is significantly longer than the migration time when larger viscosities are considered (see below).

\subsection{Disc model and viscosity prescription}\label{discmod}
For global simulations that need to be run for long times such as those we perform,  constraints 
on numerical resolution  are  such that small scale turbulence associated with angular momentum transport
has to be modelled through an effective viscosity prescription. 
 To do this  we adopt an $\alpha$-viscosity prescription (Shakura \& Sunyaev 1973).
For this the kinematic viscosity $\nu = \alpha c_s H.$
The value of $\alpha$ to be adopted depends on the expected level of turbulence
and this in turn depends on the operation of the magnetorotational instability (MRI)
(Balbus \& Hawley 1991). It is expected that there will be an inner MRI active zone 
together with an outer dead zone which may have active surface layers (Gammie 1996).
{ Small scale hydrodynamic instabilities such as the vertical shear instability may also contribute 
in the absence of the MRI (Nelson, Gressel \& Umurhan 2013).}
The location of the interface between these regions depends on details of the transition
from  magnetohydrodynamic stability  to instability  and is  subject to some degree of uncertainty.
Latter \& Balbus (2012) estimate that it  typically occurs at around 0.6 au.
{ This is illustrated in Fig. \ref{fig:DiscModel}.}

We specify the standard disc model to be such that when the unit of length used to scale our dimensionless units is 1 au
and the unit of mass is a solar mass, it corresponds to a 
steady state  model with  
accretion rate $\dot{M} = 6.0\times 10^{-10} M_{\odot}\,\mbox{yr}^{-1}.$
This is near the bottom of the range of accretion rates observed for protoplanetary discs
(Calvet et al. 2004) and might be expected to occur during their  late stages which are the focus of attention here.
However, we have also considered disc models with surface density scaled such  that  they  are up to ten times more massive and accordingly with a steady state accretion rate that is also up to ten times higher.
  
%It was assumed to be locally isothermal with constant aspect ratio $H/r =0.05.$ Thus $c_s^2 =  GM_*/(400 r).$
To allow for  an inner MRI  active  region together with an outer  region with much less activity, we specify
 $\alpha$ to be  a function of $r.$ Thus $\nu = \alpha c_sH,$
where 

%\twocolumn
\begin{equation}
\alpha = \alpha_{DZ} + \frac{\alpha_{AZ}-\alpha_{DZ}} {\displaystyle 1 + \exp\left(-25\left(1-\frac{r}{0.6}\right)\right)}.
\end{equation}
Here $\alpha_{AZ}$ corresponds to the inner region and $\alpha_{DZ}$ corresponds to the outer zone.
Our standard model has $\alpha_{AZ}=10^{-3}$ and $\alpha_{DZ}=10^{-4}.$
The functional form of $\alpha$ gives a sharp transition around $r=0.6.$
The surface density is chosen to provide the prescribed accretion rate through the relation $\dot{M} = 3\pi \nu \Sigma$.
This gives $\Sigma \propto r^{-1/2}$ when $\alpha$ is constant.

\subsubsection{Layered model} \label{layered}
\begin{figure}
\centering \includegraphics[scale=0.28]{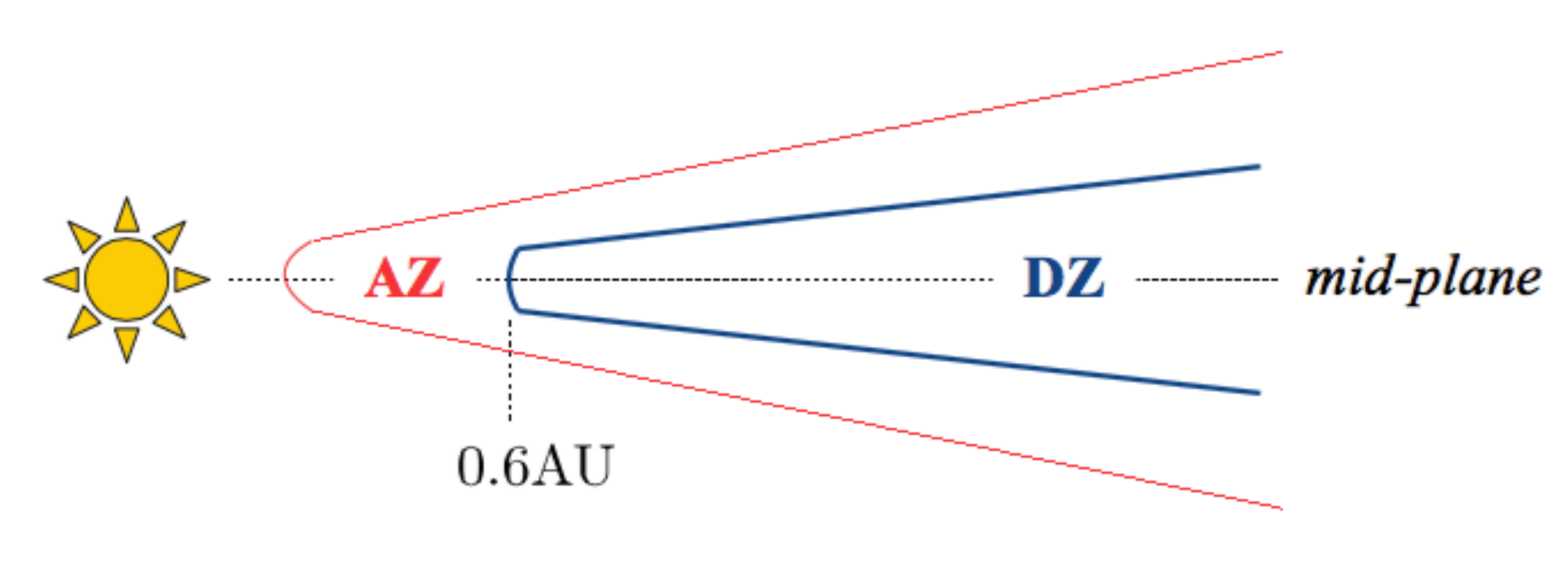}
\caption{ Illustration of our  disc model, which contains an inner MRI active zone ({AZ}) together with an outer  disc with reduced activity.
This may have a dead zone ({DZ}) together with an  active surface layer. The transition radius between those two regions is located at 0.6 au.
For standard runs $\alpha$ does not vary with $\theta$ or height in the disc.  The value of $\alpha$  then decreases sharply from $\alpha_{AZ} = 10^{-3}$ to $\alpha_{DZ} = 10^{-4}$ as the transition radius 
is passed through. For the layered model $\alpha$ is only non zero for $[\pi/2-3H/r, \pi/2] \equiv [\theta_{\mathrm{min}}, \pi/2] $ in the outer disc.}
\label{fig:DiscModel}
\end{figure}

In order to consider the possibility that a significant  effective disc viscosity arises only in the upper layers
of the outer disc (e.g. Gammie 1996) , we have  performed simulations for a layered outer disc model.
In this model  $\alpha$ was taken to be non zero only  in the upper half of the $ \theta$ domain,
namely $[\theta_{\mathrm{min}}, (\theta_{\mathrm{min}}+\pi/2)/2]$ (see Fig. \ref{fig:DiscModel}). 
The value of $\alpha$ in the upper domain was chosen to be such that the integrated stress in the meridional direction
was the same as for an outer disc model with  $\alpha$  independent of $\theta$ (see  Pierens \& Nelson 2010). 
In practice the value of $\alpha$  in the upper layers of the layered model was thus found to be $7.5$
times larger then the value for the corresponding non layered model.

For three dimensional models, the disc equilibrium pressure  is given by
\begin{equation}
P=P_0(R)\exp\left(\frac{GM_*}{c_s^2r}\ln(\sin\theta)\right).
\end{equation}
Here $R=r\sin\theta$ and the function $P_0$ can be chosen to match
a prescribed surface density or alternatively  mid  plane pressure or density.

\subsubsection{Initial gaps}\label{initgap}
In the simulations presented here the density or surface density  was modified such that  any planets
were initially placed within gaps. 
The initial orbital evolution would differ if the planets were initially embedded. However, because our goal is to measure steady state migration rates, this is not a problem.
Thus the simulations are assumed to commence after  any gaps have been formed.
 In this way  we avoid having to simulate an uncertain
 initial phase during which gaps are formed.
 
 The procedure we adopted was to reduce the density or surface density by a constant factor in the interval
 $[r_p/1.15,  1.15r_p]$ where $r_p$ is the initial orbital radius of the planet, being assumed to be in a circular orbit.   This profile was then joined to the original
 through linear connections in the intervals $[0.9 r_p/1.15, r_p/1.15]$ and $[1.15r_p,  1.265r_p].$ The gap reduction factor was taken to be a factor of ten in most cases, being a factor of one hundred for cases with planets with final masses exceeding $2$ Jupiter masses.

In addition the planets were held in fixed circular orbits for two hundred orbits before being released. Their masses
were built up to their final values over the first twenty orbits using the procedure given by de Val-Borro et al. (2007).

\subsection{Scaling to arbitrary radii}\label{scaling}
The results obtained with the unit of radius chosen to be the initial orbital radius of the inner planet can be scaled to apply to arbitrary radii.
This is done by noting that  results are invariant if the length scale is multiplied by $\lambda,$ the
time scale is multiplied by $\lambda^{3/2}$ and the mass scale left  unaltered.
To be consistent with this the surface density should be reduced by a factor $\lambda^2.$
If regions of the disc are connected in this way the situation does not correspond to
a steady state disc. However, this is not unreasonable if it is applied at radii where the age or evolution time
of the system is less than the local viscous time scale. For our standard disc model
this would be the case for length scale exceeding 2 au and age $\lesssim 2 \times 10^6\, \mbox{yr}$.

We note that when applied, the scaling procedure shifts the transition radius between the active and inactive regions while its location should in principle remain constant. However, this is not a problem as we find an insensitivity of our results to the location of the transition radius as long as the planets migrate in the outer disc (see below).

\subsection{Aspects of type II migration}\label{typeiimig}
The planets in the simulations presented here are massive enough to make  deep gaps in the disc surface density 
profile. Accordingly they undergo type II migration (e.g. Lin \& Papaloizou 1986, Lin \& Papaloizou 1993, Baruteau et al. 2014). The rate of migration is governed by
the viscous time scale and the  disc mass  within a radial scale comparable to its orbital radius, $r.$  Baruteau  et al. (2014) estimate the migration
rate of a planet of mass $M_i$  as $\tau_m^{-1},$ where
\begin{equation}
\tau_m = \tau_{\nu} {\rm max} \left(1, \frac{M_i}{4\pi\Sigma r^2}\right), \label{typ2}
\end{equation}
where $\tau_{\nu}$ is a viscous time scale. Thus when
the planet mass becomes large the evolution rate slows on account of the inertia of the planet.
Note that  Ivanov et al. (1999) obtain a faster rate than  implied by equation (\ref{typ2}),  finding that the second term in the brackets appears taken to a fractional power.
The rate is faster  because disc material tends to pile up near the outer gap edge increasing the angular momentum flux that the planet needs to provide in order to maintain it.

\section{Numerical results} \label{Numres}

\subsection{Comparison with previous results} \label{CompPreviousResults}
{ 
Because  our simulations of planet-disc interaction with the \texttt{PLUTO} code were implemented from scratch, we begin by establishing that some of the main  results  obtained in previous studies are recovered  with our code.  In particular  we focus on the much studied GJ 876 system (see e.g. Snellgrove et al. 2001, Kley et al. 2005, Crida et al. 2008), as well as a case involving the action of the so called Masset-Snellgrove mechanism invoked to reverse type II migration (Masset \& Snellgrove 2001) that has been considered by  Crida et al. (2009b).}

\subsubsection{The case of GJ 876}\label{GJ876}
{ \begin{figure}
\centering \includegraphics[width=7.5cm]{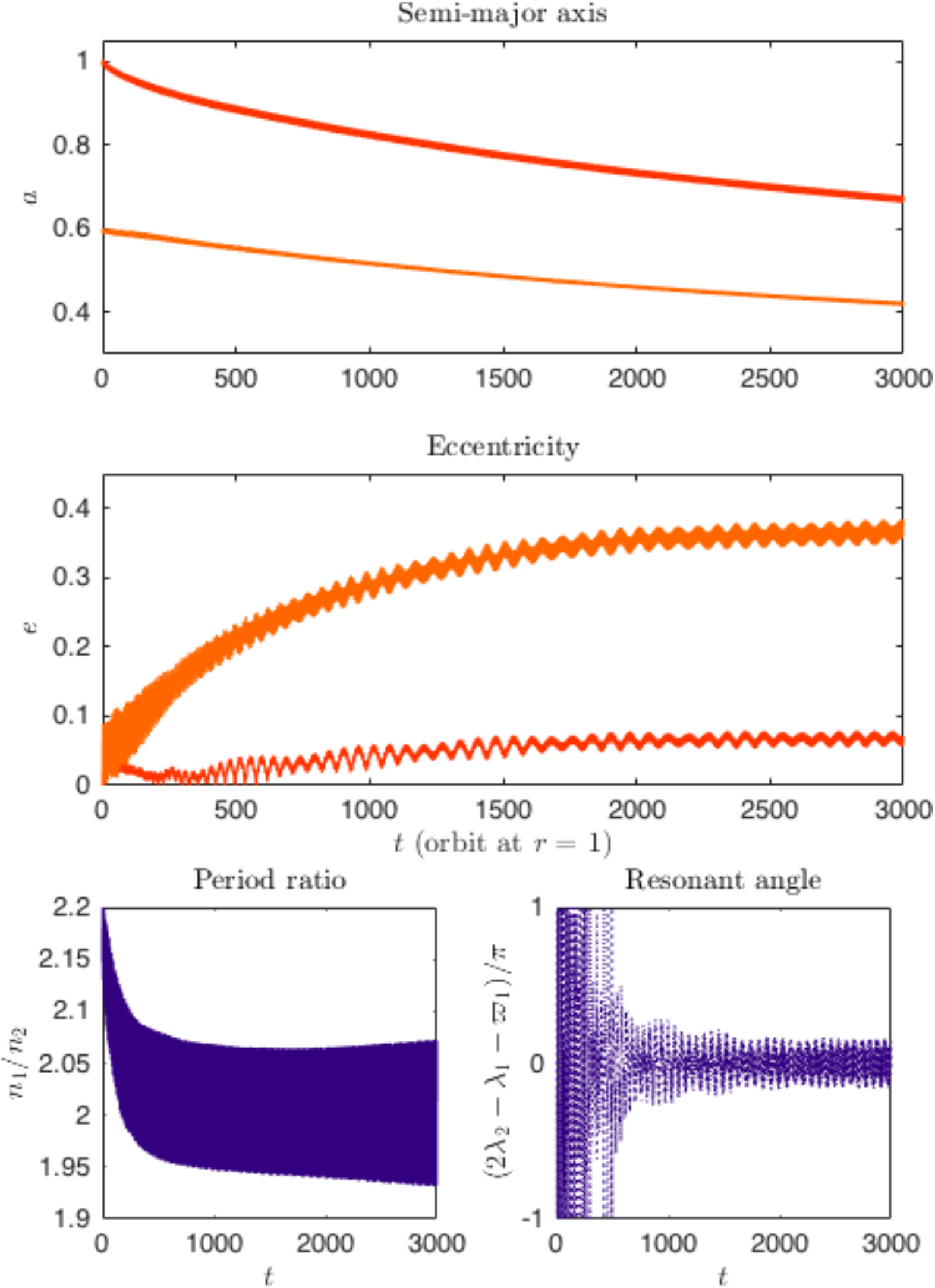}
\caption{{ Results from the GJ 876 comparison run.
The uppermost panel shows the evolution of the semi-major axes of the two planets,
the middle panel shows the eccentricities (the upper curve corresponding to the inner planet), the left hand  bottom panel shows the
period ratio for the two planets  and the right hand bottom panel shows  the  resonant angle  $2\lambda_2  - \lambda_1 -\varpi_1.$
The unit of time, $t,$  in this and subsequent figures
 is the orbital period at a value of the dimensionless radius equal to unity.}}
\label{fig:GJ876}
\end{figure}

Our first comparison run was initiated with the mass ratio  equal to $6 \times 10^{-3}$ for the inner planet, and $1.8 \times 10^{-3}$ for the outer planet. These parameters correspond to those of GJ 876. The simulation  employed the physical setup described in Snellgrove et al. 2001 (see their Section 3.1).  A  constant aspect ratio $h = 0.07$, and a constant $\alpha$-viscosity prescription with $\alpha = 2\times 10^{-3}$ are adopted. The two planets are initially in circular orbits, and start their evolution with semi-major axis $a_1 = 1.0$ for the outer planet, and $a_2 = 0.6$ for the inner planet.   Note that   for this comparison,  we adopt
their nomenclature and system of 
dimensionless units. Hence, the outer planet is  initially located outside the exact 2:1 commensurability ($a_1 \sim 0.95$).

A putative uniform  initial disc  surface density $\Sigma_0$  corresponds to what  would give a disc mass of $2M_J$ within the initial orbit of the outer planet. It is also assumed that both planets are located inside a tidally truncated cavity located at $r < 1.3$, with low surface density equal to $0.01 \Sigma_0$. In the region $1.3 < r < 1.5$, the surface density is prescribed such that $\ln \Sigma$ linearly joins to $\ln \Sigma_0.$
 In addition in our run, the smoothing filter of Crida et al. (2009a) was used when calculating torques acting on planets.

The results are displayed in Fig. \ref{fig:GJ876} which  can be compared  with Fig. 1 of Snellgrove et al. (2001). The two planets first  undergo a phase of convergent migration. At time $t \sim 400$ orbits, the period ratio between the two planets locks around the value 2, and the resonant angle $2\lambda_2  - \lambda_1 -\varpi_1$ starts to librate around 0.
Here $\lambda_2,$ $\lambda_1,$ $\varpi_2$ and $\varpi_1$ are the longitudes of the inner and outer planet
 and the longitudes of pericentre for the inner and outer planets respectively.
 A 2:1 mean motion resonance is subsequently maintained.
The behaviour we obtain is very similar to that found by Snellgrove et al. (2001)
until a  run time  1500 orbits is reached. 
After this the evolution of the  semi-major axes and eccentricites stall in their run while they   continue to  respectively decrease and increase slightly in ours.
We believe that the stalling  is artificial and due to the approach of the inner planet to the inner boundary,
located at a radius equal to 0.4 in their case.  In order to avoid this,  we chose an inner boundary radius $r_{\mathrm{in}} = 0.2$.
This allowed us to continue the evolution of the two planets  for up to  3000 orbits without  this type of influence from the inner boundary.
In the context of the above discussion,  we remark that Kley et al. (2005) performed simulations with the inner planet totally interior to the calculation domain.
They found that its eccentricity continued to increase as we did (see their Fig. 10).
%The issue of continued eccentricity growth was also later addressed by Crida et al. (2008), who invoked the interaction with an  inner disc in order to prevent it. 

\subsubsection{An example illustrating the Masset-Snellgrove mechanism }\label{MSmech}
\begin{figure}
\centering \includegraphics[width=7.5cm]{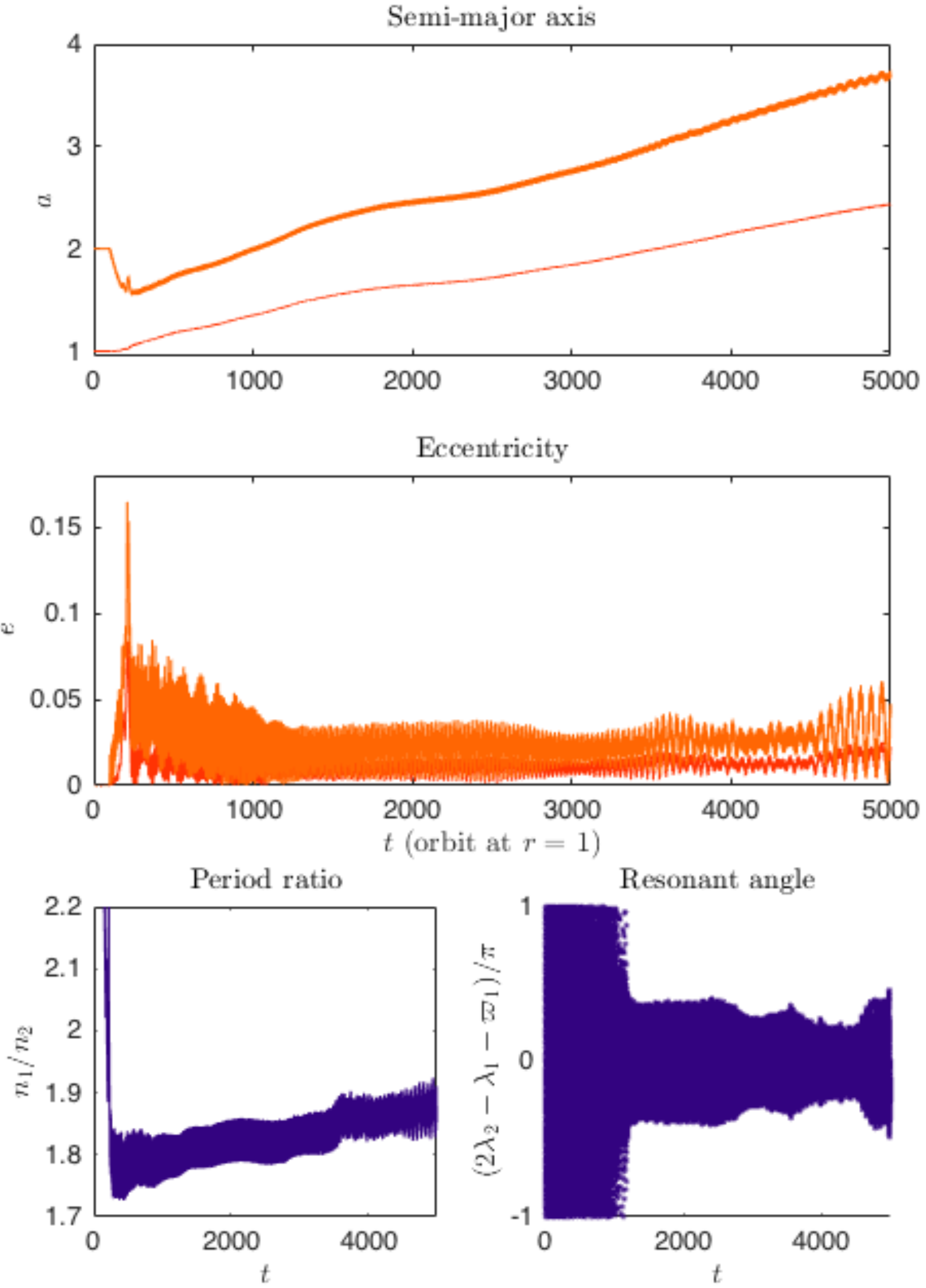}
\caption{{ Results from the run illustrating the Masset-Snellgrove mechanism. The panels correspond to those of Fig. \ref{fig:GJ876}.}}
\label{fig:Crida09}
\end{figure}

Masset \& Snellgrove (2001) found that the migration of two planets locked in a mean motion resonance can proceed outwards.  For this to occur they must open overlapping gaps and the outer planet be of significantly lower mass  than the inner one. Our second comparison run was chosen such that  this  mechanism is expected to operate. It was initiated with mass ratios for the inner and outer planets respectively  equal to $3 \times 10^{-3}$, and $10^{-3}.$  These parameters correspond to those adopted by Crida et al. (2009b).
%who invoke the Masset-Snellgrove mechanism to account for the presence of giant planets up to 120 au to their host stars that have been observed by direct imaging (e.g. Kalas et al. 2008, Marois et al. 2008).
 The simulation adopted their physical  setup  apart from  the treatment of the boundaries  (see their Section 3.2).
 They employ   an  additional  matched  one dimensional simulation of a  putative  enveloping  disc, whereas we adopt our standard  conditions described above.
  As the inner boundary radius is located at  a radius equal to $45\%$
of the initial inner planet orbital radius and the interior disc plays a major role in driving the outward migration, we  expect differences in  results  at early times.
The outer boundary may also become significant at late times.
The form of the  aspect ratio adopted is  $h = 0.045 \times (r/a_0)^{1/4}$ ($a_0$ being the initial inner planet orbital radius), the initial surface density
is $\Sigma(r) = \Sigma_0 \times (r/a_0)^{-3/2}$ with $\Sigma_0 = 1.5 \times 10^{-3}$, and an  $\alpha$-viscosity prescription with $\alpha =0.01$ is adopted.
The two planets are held in circular orbits for the first 100 orbits of the inner planet, and start their evolution with semi-major axis $a_1 = 1 = a_0$ for the inner planet, and $a_2 = 2.0$ for the outer planet.
 % the units  adopted such that $a_0 =1.$
%In addition, the smoothing filter of Crida et al. (2009a) was used when calculating torques acting on planets.

The results are displayed in Fig. \ref{fig:Crida09} which can be compared  with Fig. 1 of Crida et al. (2009b).
 After the release, both planets start to  migrate inwards.  The outer planet moves rapidly in a type III migration regime with  the inner one moving
 much more slowly corresponding to type II migration.
 This  convergent migration of the two planets causes  passage  through the 2:1 mean motion resonance.
  The resonant angle $2\lambda_2 - \lambda_1 - \varpi_1$ starts librating around zero from time $t \sim 1000$. 
  Subsequently, the convergent migration stops and the two planets migrate smoothly outwards together.

The qualitative behaviour described above is very similar to that obtained by Crida et al. (2009b).
However,   they  obtain  an acceleration at early times followed by a significant  slow down later on,  whereas in our case the  outward migration rate is more uniform.
 As indicated above, this difference is not unexpected on account of the role of
the inner boundary.  Thus we find that just after the planets start moving outwards,  the inner planet initially migrates outwards  at half the rate obtained by Crida et al. (2009b). However, after 
5000 orbits at $r=a_0, $ (corresponding  to a time $t \sim 1.12 \times 10^5\, \mathrm{yrs}$ in Crida et al. 2009b), when the inner boundary is expected to be less important,
 the mean migration  rates are approximately the same.

   % The fact that they use the 2D1D version of \texttt{FARGO} (Masset 2000), in which the standard two-dimensional polar grid is surrounded by a one dimensional grid to compute the whole %physical extension of the disc (Crida et al. 2007) could account for this difference.

%It is worth mentioning that the period ratio behaviour is somewhat unexpected. Indeed, although the 2:1 resonant angles are librating,  indicating the importance of the 2:1 mean motion resonance, the average ratio between the orbital periods of the two planets locks at $\sim 1.8$ before rising slowly and smoothly up to $\sim 1.85$ by the time $t \sim 5000$ (see Fig. \ref{fig:Crida09}). We note that Crida et al. (2009b) obtain similar period ratio values.

%As an illustration of the behaviour described in the above discussion, Baruteau \& Papaloizou (2013) found that the interactions of a planet with the wake of its companion can lead to such  divergent migration. They found that this wake-planet interaction is particularly efficient when at least one of the planets opens a partial gap around its orbit. In our run, this is the case of the outer planet. 
} %end of bold font due to new section 4.1
%end  of section 4.1
 
\subsection{The standard run}\label{stdrun}
{ We now consider the runs that are the focus of this paper
which, unlike in Section \ref{MSmech}  involve for the most part inward convergent migration of the planets.
For these cases as well as that of GJ876 an overview can be obtained by considering a simple
$N$ body model in  which the planets move as  particles under their gravitational interaction and the influence of additional forces presumed
to arise from interacting with the disc  (see eg. Snellgrove et al. 2001, Lee \& Peale 2002, Nelson \& Papaloizou 2002).
These result in orbital circularization and  migration. It is found that unless the convergent migration is very rapid,
the planets attain a commensurability and then migrate together maintaining it. As they do so their eccentricities  increase until either their migration halts or  their
rate of growth can be balanced by a damping  process.

 For the slowest migration rates a 2:1 resonance is attained
for Jovian mass planets while for larger masses a 3:1 resonance can be  attained. As the rate of convergence is increased
closer commensurabilities are attained. Our results are fully in line with these general expectations.
The rate of convergent migration and hence the closeness of the commensurability 
is determined by the rate of angular momentum transport in the disc which for a fixed $\alpha$ distribution
increases with the mass in the disc. }

\begin{figure}
\centering \includegraphics[width=7.5cm]{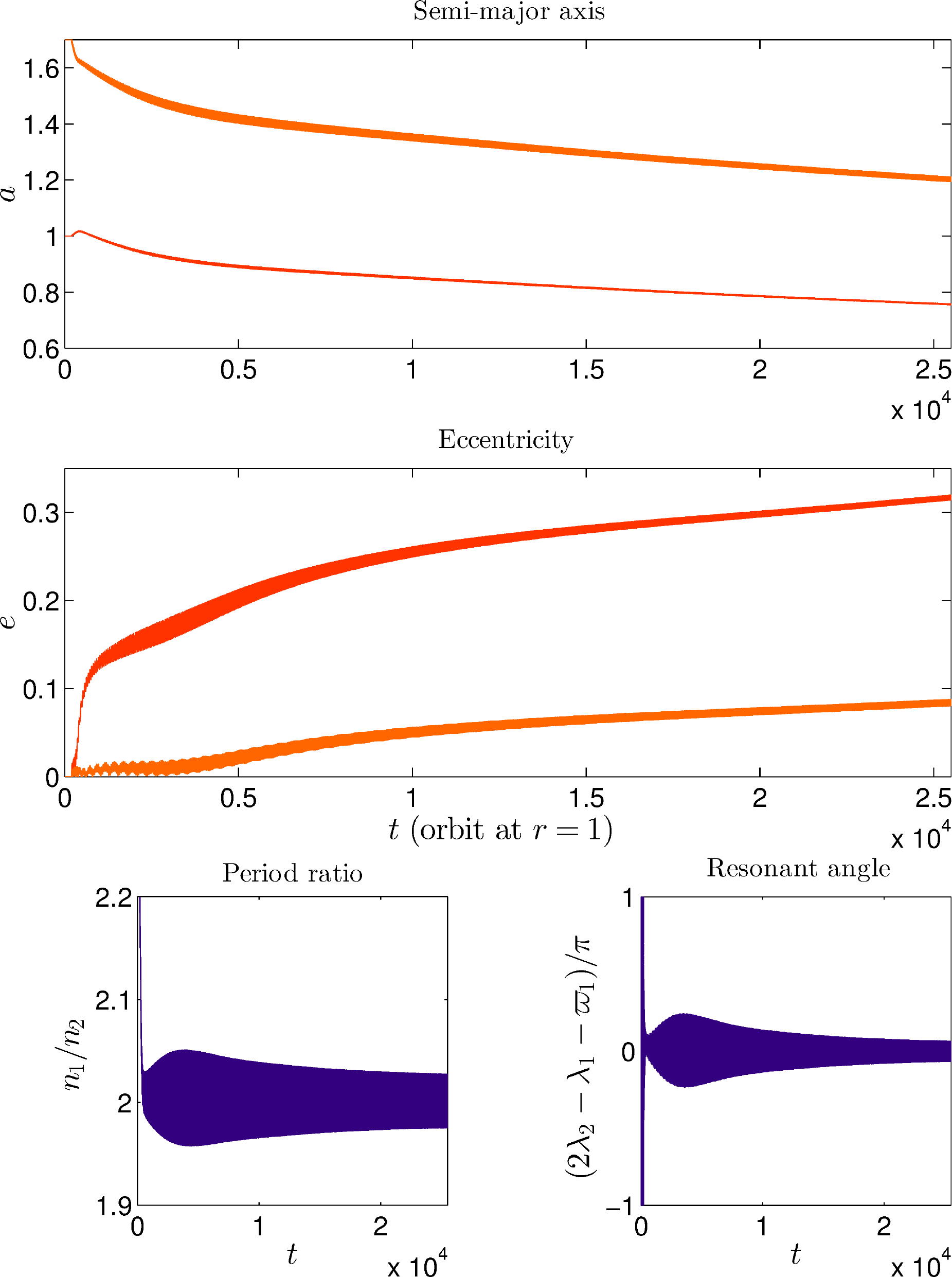}
\caption{Results for the 2D  standard run. { The panels correspond to those of Fig. \ref{fig:GJ876}.}
}\label{fig8}
\end{figure}

The 2D standard run was initiated with the mass ratio for both planets equal to $10^{-3}.$
Noting that they  are somewhat uncertain,   these parameters  may  approximately correspond  to 
 those for HD 155358 and  24 Sextantis (see  table \ref{table1}).
The simulation  was initiated with the planets occupying a common gap in the standard disc as described above. We assume they start their evolution with semi-major axis $a_1 = 1$ for the innermost planet, and $a_2 = 1.7$ for the outermost planet in dimensionless units.
 The results  are plotted in Fig. \ref{fig8}.
The  evolution of the semi-major axes and eccentricities are shown for a time interval of $2.5 \times 10^4$ time units.
%The unit of time in this and subsequent figures
% is the orbital period at a value of the dimensionless radius equal to unity or when this corresponds to 1 au.
The system enters 2:1 resonance after a few hundred orbits.
Note that in this and other cases,  the inner planet  migrates outwards at early times on account of the influence of the inner disc.
The lowermost left hand panel shows the evolution of the orbital period ratio $n_1/n_2$
of the two planets, $n_1$ and $n_2$ denoting the mean motions of the inner and outer planet respectively. An ultimate libration amplitude of a few percent is indicated. 
The lowermost right hand panel of Fig. \ref{fig8}  shows the
 resonance  angle,  $2\lambda_2-\lambda_1-\varpi_1,$ appropriate for  the 2:1 resonance. 
% Here $\lambda_2,$ $\lambda_1,$ $\varpi_2$ and $\varpi_1$ are the longitudes of the inner and outer planet
 %and the longitudes of pericentre for the inner and outer planets respectively. 
 This resonance angle ultimately librates around zero.
 The behaviour of the second resonance angle, 
$2\lambda_2- \lambda_1-\varpi_2,$  is very similar.
 
\begin{figure}
\centering
\includegraphics[height=1.8in,angle=0]{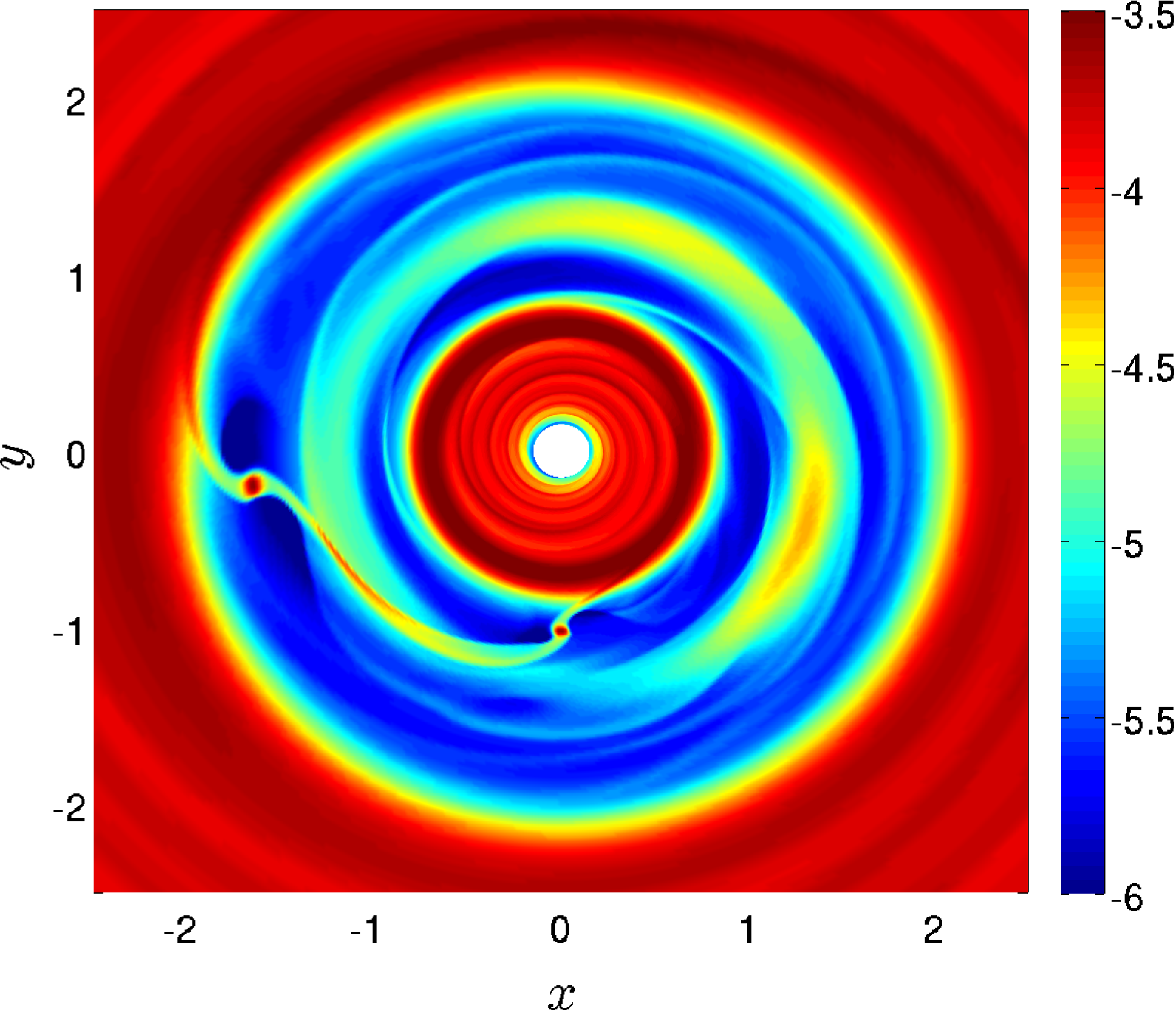}\vspace{0.2cm}
\includegraphics[height=1.7in,angle=0]{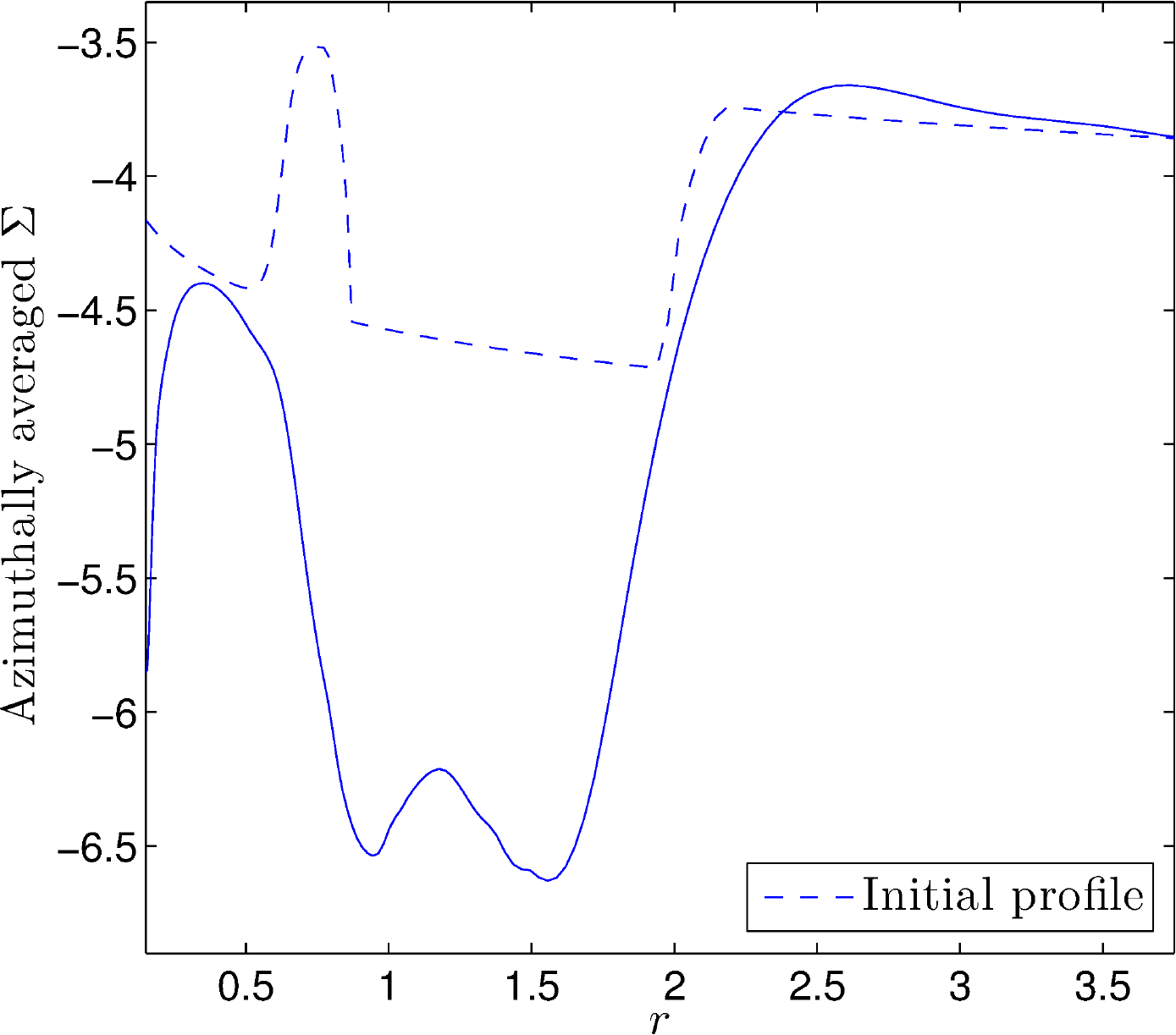}
\caption{The upper  panel shows surface density  contours in logarithmic scale for the disc
 after $280$ orbits  of the simulation shown in Fig. \ref {fig8}.  The lower panel shows the  azimuthally 
averaged surface density  in logarithmic scale after $3750$ orbits (solid curve) and the initial surface density (dashed curve).
The planets occupy the deep common gap.}
\label{fig11}
\end{figure}

 At late times, the  migration time   $-r/{\dot r} \sim 1.5\times 10^5$ inner planet initial orbital periods.
 For comparison the viscous time  $2r^2/(3\nu)$ for the outer disc is $4.2\times 10^5$ orbits at $r=1$ in dimensionless units.
 Note that the two resonantly coupled  planets migrate at a similar rate as  a single planet (see below). 
 %Adopting  the scaling procedure outlined in section \ref{scaling}
% we can estimate the radius the inner planet   started at if the system migrated  in resonance for a time, $t,$ in order to get to its final location,
%assumed to be at a significantly smaller radius,   as
%$r  \sim  ( t/( 1.5\times 10^5\, \mbox{yr}))^{2/3}\, \mbox{au}    \sim 3.5\, \mbox{au}$ for $t= 10^6\, \mbox{yr}.$ 
 Additional results from  the 2D standard run are illustrated in Fig. \ref{fig11}.
  The upper panel shows surface density  contours for the disc
 after $280$ orbits and the lower panel shows the  azimuthally
averaged surface density gap after $3750$ orbits.
Note the non axisymmetric vortex like structures in the low surface density ring between the planets { (see e.g. de Val-Borro et al. 2007)}.
The planets  occupy a deep common gap that is characteristic of the simulations reported here.
 
\begin{figure}
\centering \includegraphics[width=7.5cm]{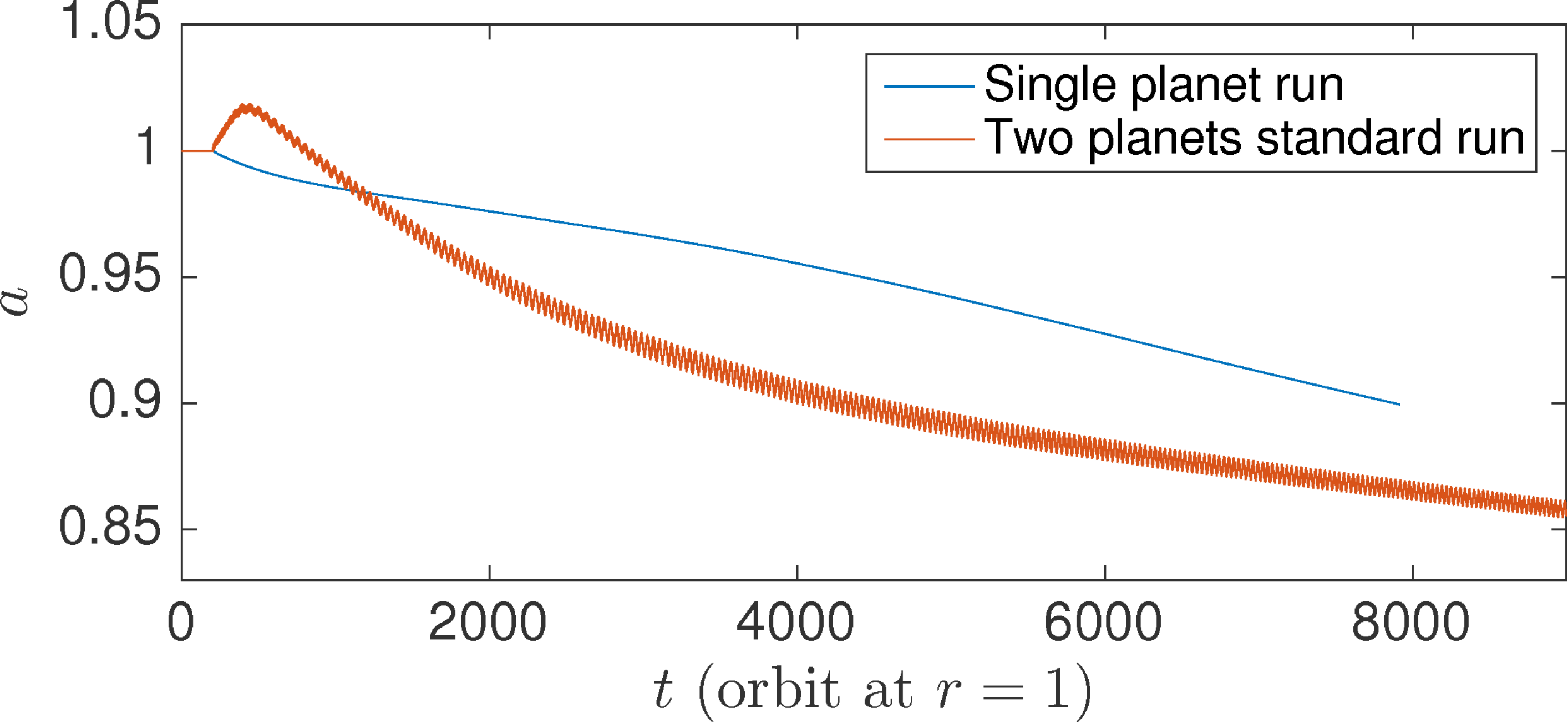}%{Comp_SingleTwoPlanetsA-eps-converted-to.pdf}
\caption{{ Evolution of the semi-major  axis  of a single planet  with mass ratio $10^{-3}$ migrating in a standard disc 
plotted together with the evolution of the semi-major axis for the innermost planet in the standard run.
The evolution is shown as a function of time expressed in units of the  orbital period at $r=1$ in dimensionless units.}
\label{figsingle}}
\end{figure}

 In Fig. \ref{figsingle} we illustrate  the migration of a single planet in a standard disc.
 { For comparison the evolution  is plotted together with that for the inner planet in the standard case.
 At the beginning the evolution in the standard case is outwards. This is because the planet starts in a much wider gap
 and is pushed outwards by the inner disc
 in that case. However, once the system attains resonance the planet is pushed inwards by the inwardly migrating outer planet.
 For the isolated planet, the inward migration is driven by the outer disc directly. In both cases we expect a characteristic migration rate
 corresponding to type II migration and indeed the rates are found to be ultimately  comparable.
 The migration speed  of the single migrating  planet   attains a steady  value  between $5000$ and $8000$ orbits. The  mean  migration
  time scale  over this period is estimated as
 $\langle-r/ {\dot r}\rangle \sim 6.9 \times 10^4$ orbits at a dimensionless radius of unity.  
This is characteristic of type II migration, an aspect that is discussed further in Section \ref{Disc}.}

% Assuming a central  solar mass
% and the scaling described in section \ref{scaling} applies, the radius of the inner planet at which its migration has to start, 
% given it takes a time, $t,$ to arrive at its final location   can be estimated as
% $r = ( t/( 1.1\times 10^5\, \mbox{yr}))^{2/3}\, \mbox{au}  \sim 4.4\, \mbox{au}$ for $t= 10^6\, \mbox{yr}.$  

\subsubsection{An entirely active disc}
 Fig. \ref{fig4} shows  results for a simulation with the same  conditions,  including the
 initial set up of the pair of planets,
as for the 2D standard run,  except that only an active model disc was used.
That is the transition radius of $r=0.6$ for the standard run was  effectively moved to very  large radii. 
The quantities plotted in the various panels of Fig.  \ref{fig4} correspond to those  plotted in  the corresponding panels of Fig. \ref{fig8}
 and the results are qualitatively very similar.
The estimated  late time  migration  time $-r/{\dot r} \sim 2.4\times 10^5  $ orbits  is significantly slower  than in  both the standard two planet   case and the single planet case 
where the planet migrates in the inactive region. This is because even though the viscosity is ten times larger in the active disc, the surface density is  ten times smaller
resulting in a slower migration rate on account of the reduced disc mass in the neighbourhood of the planets.

\subsubsection{A disc without an inner active region}
\begin{figure}
\centering \includegraphics[width=7.5cm]{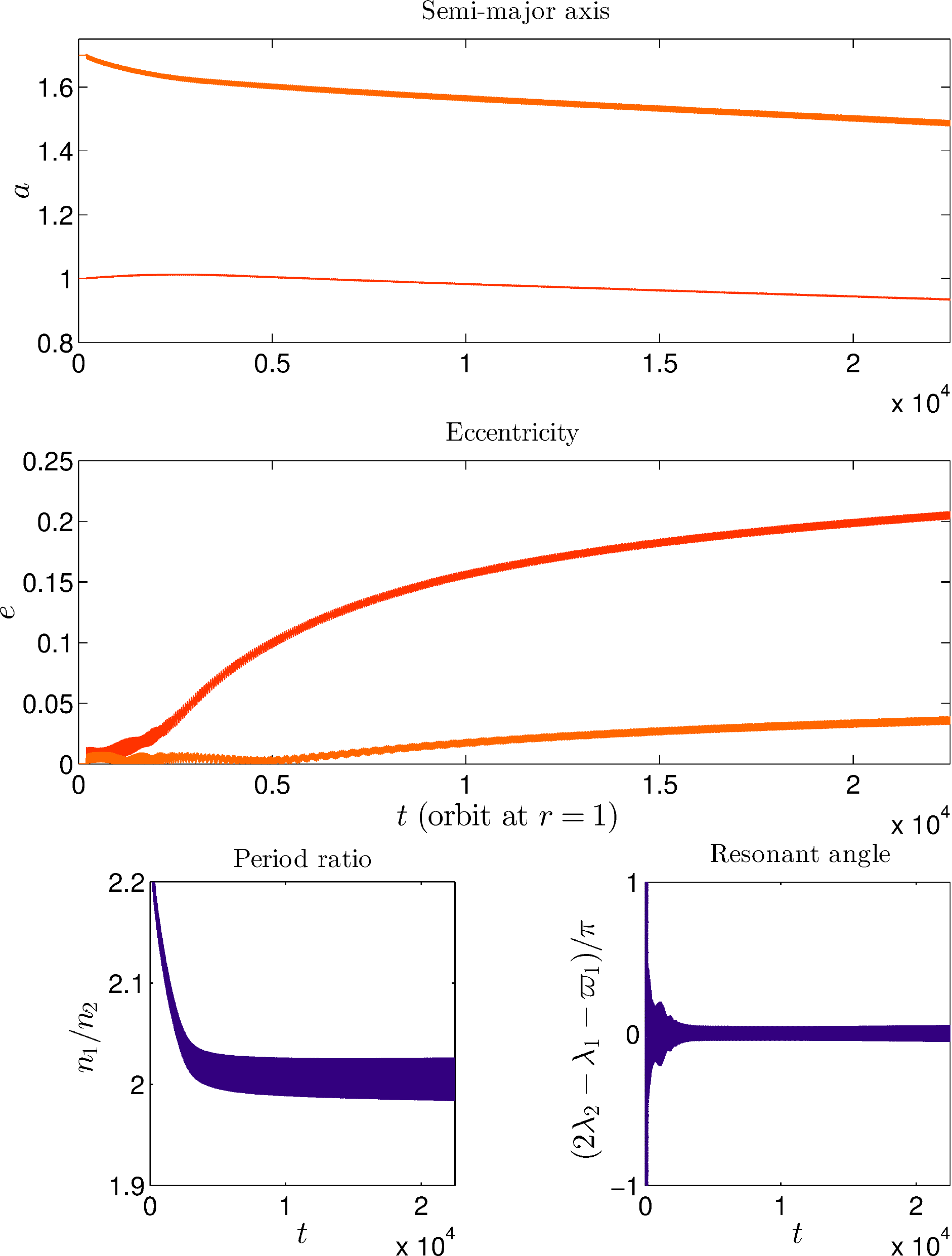}
\caption{  
Results of a simulation with  the same  conditions
as for the 2D standard run except that only an active model disc was used.
That is the transition radius of $r=0.6$ for the standard run  can be regarded as being moved to a very  large value. The panels correspond to those of Fig. \ref{fig8}.
Note that the migration  is significantly slower  than in the standard  case. }\label{fig4}
\centering \includegraphics[width=7.5cm]{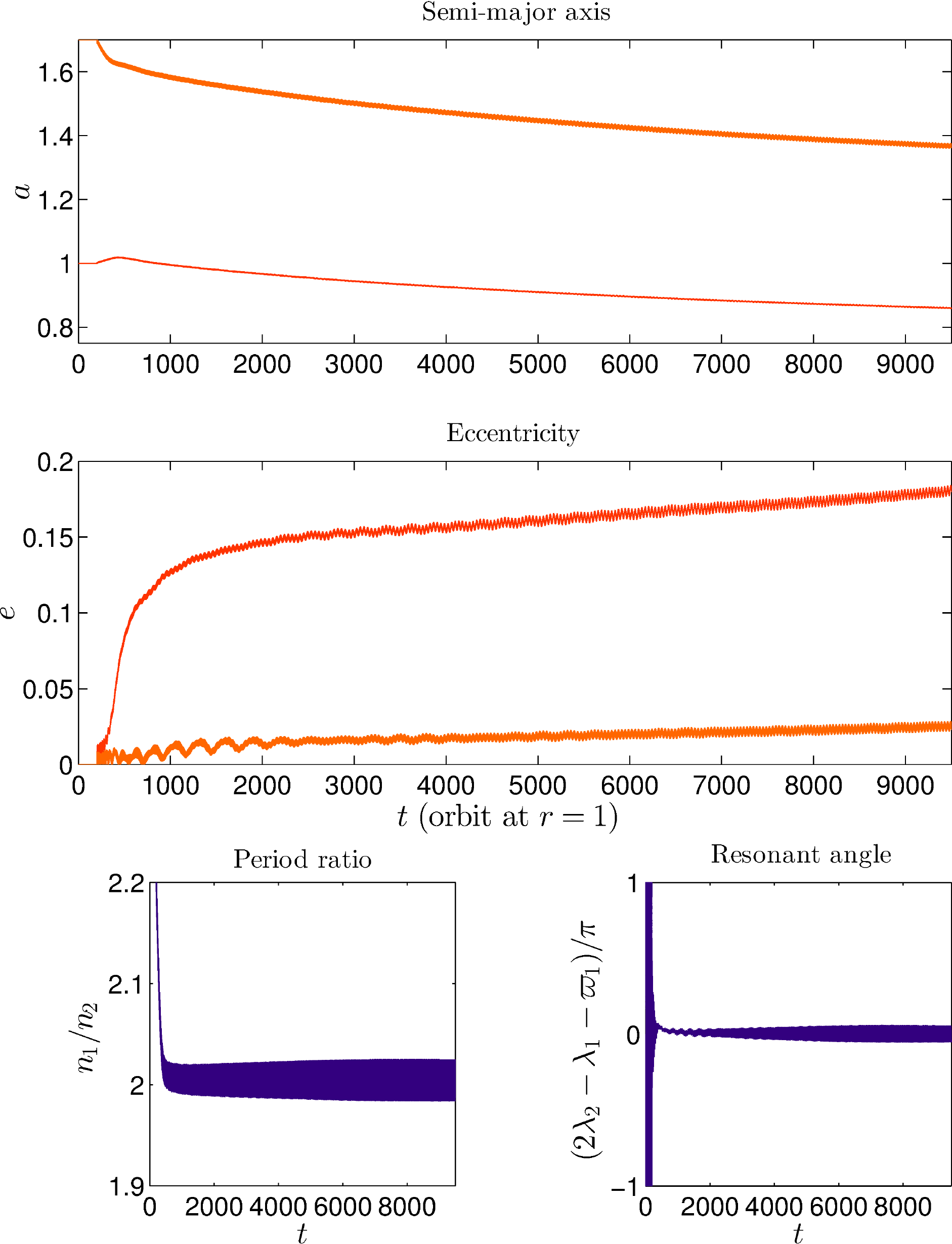}
\caption{As for  the 2D standard run shown in  Fig. \ref{fig8} except that  in this case  the disc model 
was  such that the inner active disc
was removed or equivalently the transition radius can be regarded as being  moved to a very small value. The panels correspond to those of Fig. \ref{fig8}.
 }\label{fig5}
\end{figure}

 In Fig. \ref{fig5}  we  illustrate a 2D two planet run  with the same parameters as the standard run except that the inner active disc
was removed or equivalently the transition radius was moved to very small values.
The quantities shown in the  panels of Fig. \ref{fig5}  are the same as in the corresponding panels of Fig. \ref{fig8}. 
The migration time scale at later times 
is estimated  to be $-r/ {\dot r} \sim   10^5$ orbits at a dimensionless radius of unity.
This is almost exactly the same as for the standard case at the same time.
%The radius of the inner planet,  at which its migration  
% in resonance starts,  given it takes a time , $t,$ to arrive at  its final location,  can be estimated as
% $r =  (t /(  10^5 y.))^{2/3}au  \sim 4.6 au$ for $t= 10^6 y.$  
Note that the eccentricities at times corresponding to the same amount of relative joint migration  are smaller in this case  than those for the corresponding standard case, 
which is indicative of a larger damping rate.

\begin{figure}
\centering
\includegraphics[width=7.5cm]{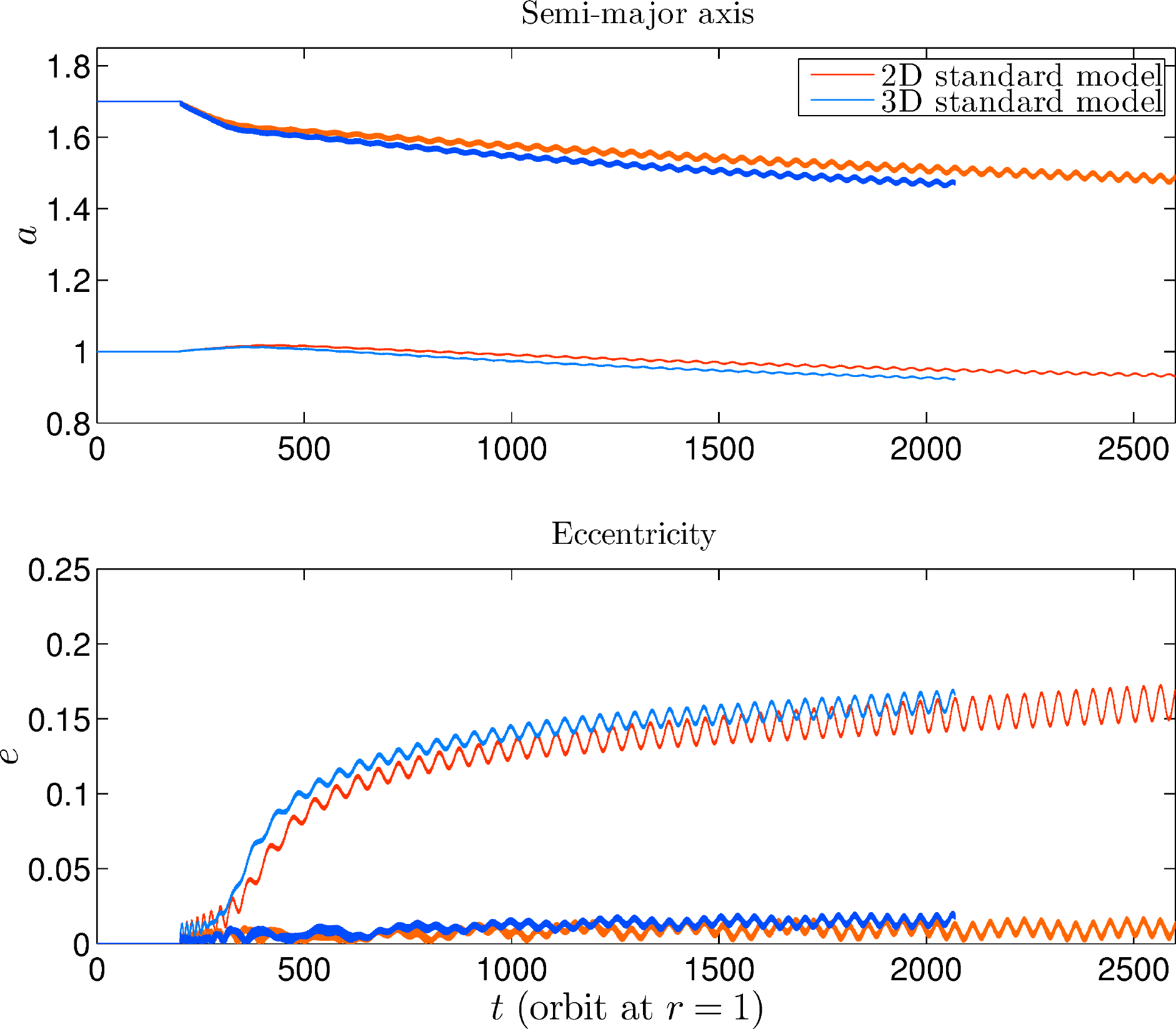}
\caption{Comparison between the  2D and  3D  standard runs.
A comparison of the evolution of the  semi-major axes
is  given in the upper panel. The lower panel shows the
evolution of the eccentricities with the upper pair of curves corresponding to the inner planet.}
\label{fig24}
\end{figure}
%\onecolumn
\begin{figure*}
\centering
\includegraphics[width=17cm]{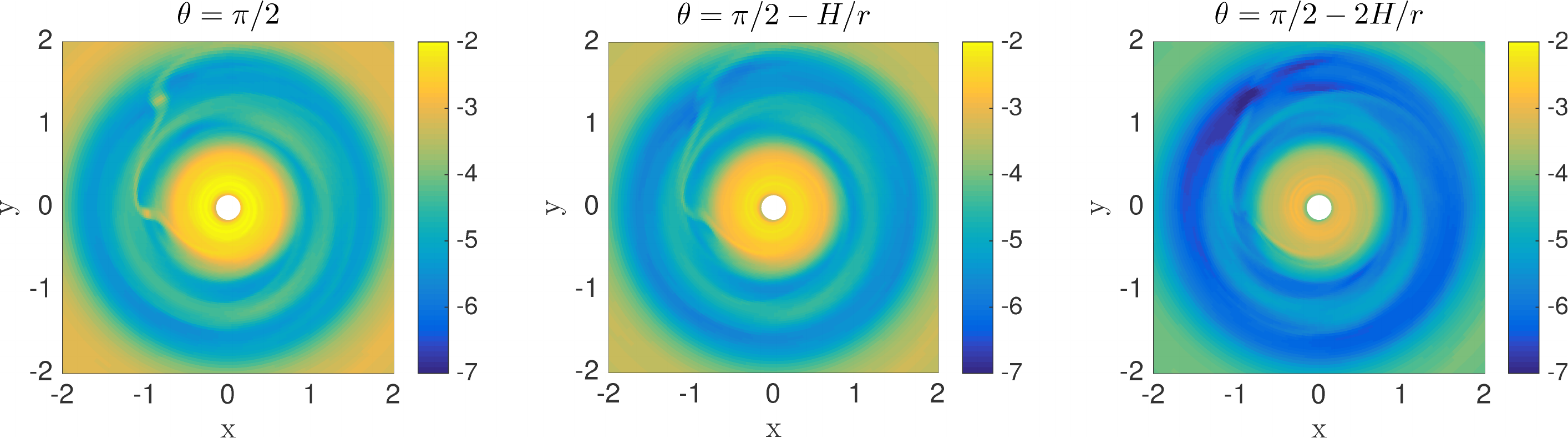}
\caption{ {  Contours of $\log \rho$ in dimensionless units in the $(r,\varphi)$ plane  are shown  for three indicated values of $\theta$
for the 3D standard run after $1100$ orbits.
From left to right these values correspond to the mid plane and approximately to heights, $H$ and $2H$ above the mid plane.}
}
\label{fig24a}
\end{figure*}

\begin{figure}
\centering
\includegraphics[width=8cm]{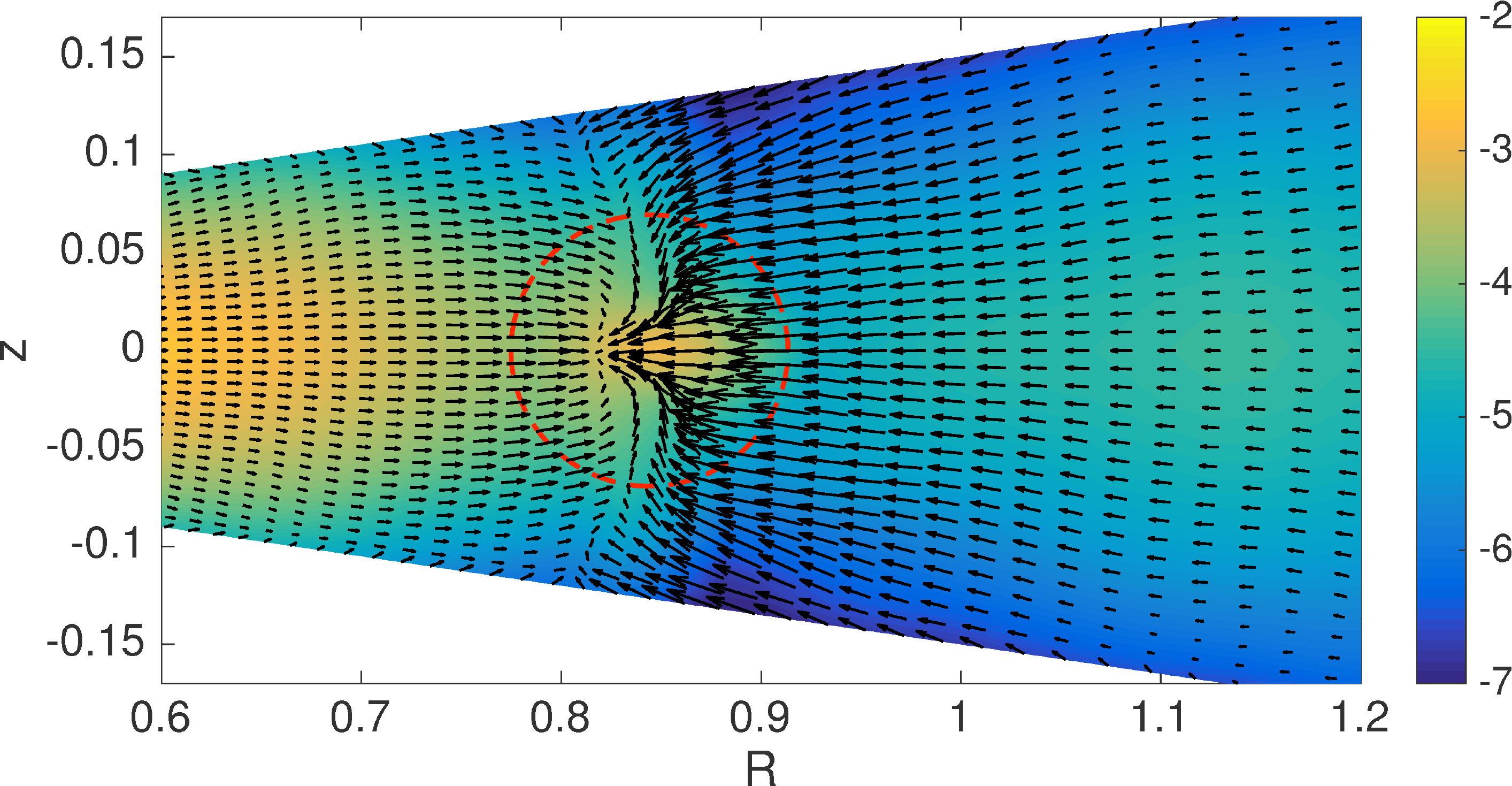}%{VertCut800.eps}
\caption{Stream lines in a meridional section
 at the azimuth of the location of the inner planet for the 3D standard run  after  $1410$ orbits.
{ The colour scale indicates $\log \rho$ in dimensionless units and the dashed circle indicates the location of the surface of the Hill sphere.} }
 \label{fig28}
\end{figure}
\subsubsection{Three dimensional simulations}
{ We now consider 3D simulations. The  disc models were set up as described in Sections \ref{discmod} and \ref{layered}
with the planets introduced together with inital gaps as described in Section \ref{initgap}. We recall that
for the standard 3D model $\alpha$ does not vary with $\theta.$}

 A comparison of the evolution of the semi-major axes
for the standard 2D  two planet with the  corresponding results from  the 
standard 3D run is given in the upper panel of Fig. \ref{fig24}.
The evolution of the eccentricities is shown in the lower panel. 
These are found to be in very good agreement { with each other and therefore also in accord
with the simplified $N$ body approach mentioned above.}
{ This suggests  that the evolution can be determined
by considering the 2D response of the vertically averaged disc.
The fact that the response is for the most part two dimensional
is indicated by the behaviour of the disc state variables.
We show  density  contours in logarithmic scale  for three indicated values of $\theta$
at a typical late time of  $1100$ orbits after the start of the simulation in Fig. \ref{fig24a}.
These values correspond to the mid plane and approximately to  heights, $H$ and $2H$ above it. 
It will be seen that apart from in the neighbourhoods of the planets, the density distributions are
approximately the same at the different heights apart from a constant scaling factor.
In the neighbourhoods of the planets there are  the local mass concentrations usually seen 
in 2D simulations (see left panel of Fig. \ref{fig24a}). In the upper regions of the disc these are absent and there is instead material depletion on account of  vertical flows towards the planets. To illustrate this aspect}
Fig. \ref{fig28} shows the charactersitic form of the  stream lines in a meridional section
 at the azimuth of the inner planet. These are shown after a time corresponding to  $1410$ orbits at dimensionless radius equal to unity.
Significant vertical motions associated with material moving from the upper regions of
the disc towards the mid-plane slightly interior to the location of the centre of gravity of the planet  are apparent. This is an  effect that can only
be represented in 3D. However, this is not  found to cause significant departures from the 2D results for the orbital evolution,
presumably because of the small amount of material in the gap regions near to the planets.
{ We remark that a similar behaviour of the state variables  to that described here was found by Pierens \& Nelson  (2010)
in their 3D  simulations with a single planet.}

\subsubsection*{Layered model}\label{Layered}
\begin{figure}
\centering
\includegraphics[width=7.5cm]{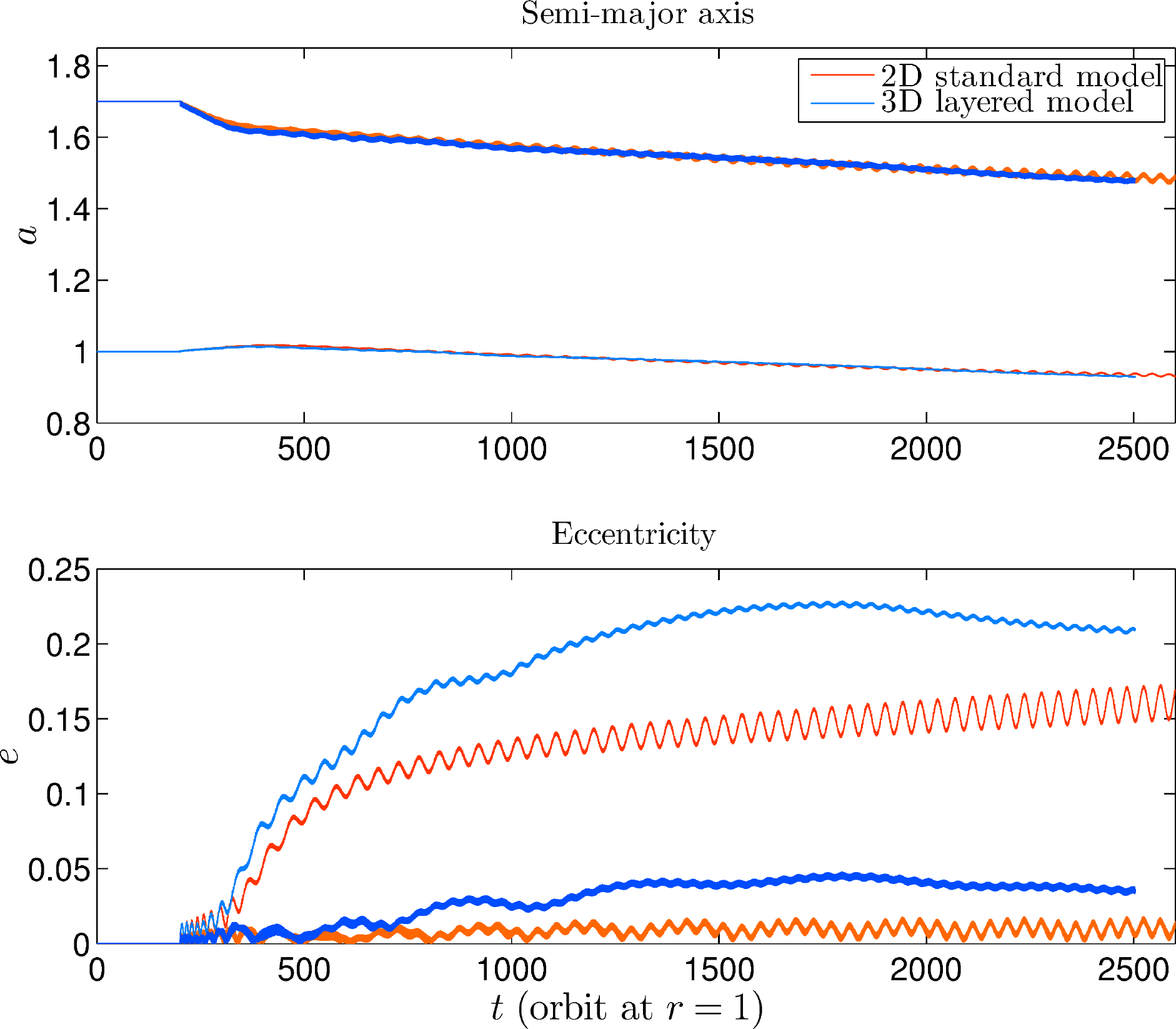}
%\caption{ Comparison between 2D and  3D  standard two planet run}\label{fig28}
%\includegraphics[height=2.0in,angle=0]{11-11.7-1-101-e1.eps}
%\includegraphics[height=2.0in,angle=0]{11-11.7-1-101-flow800.eps}
\caption{Results for  the 3D layered model compared to those of the 2D standard run.  The panels correspond to those of Fig. \ref{fig24}. Note that the lower panel shows the
evolution of the eccentricities with the uppermost pair of curves corresponding to the inner planet.}\label{fig24l}
\end{figure}

A comparison of the results  obtained for the layered model described in Section
\ref{layered}  with those obtained from the standard 2D model are illustrated in Fig. \ref{fig24l}.
The layered disc model had the same integrated stress in the meridional direction as in the standard  case.
Apart from the differing disc model, the conditions are the same as for the standard 3D run.
It will be seen that the evolution of the semi-major axes  for the two runs is almost identical.
{ We remark that the migration rates for both planets are slightly faster in the standard case 
as compared to the simulation with the layered model. This is in line with the results of  Pierens \& Nelson (2010)
for the case of a single Jupiter mass planet.}

However, the eccentricities are significantly larger at corresponding times  for the layered model.
This indicates that the eccentricity damping rate is lower in this case and accordingly it depends on the
detailed properties of the disc model. { For the layered model, we recall that the disc is inviscid near the mid plane.}

\begin{figure}
\centering \includegraphics[width=7.5cm]{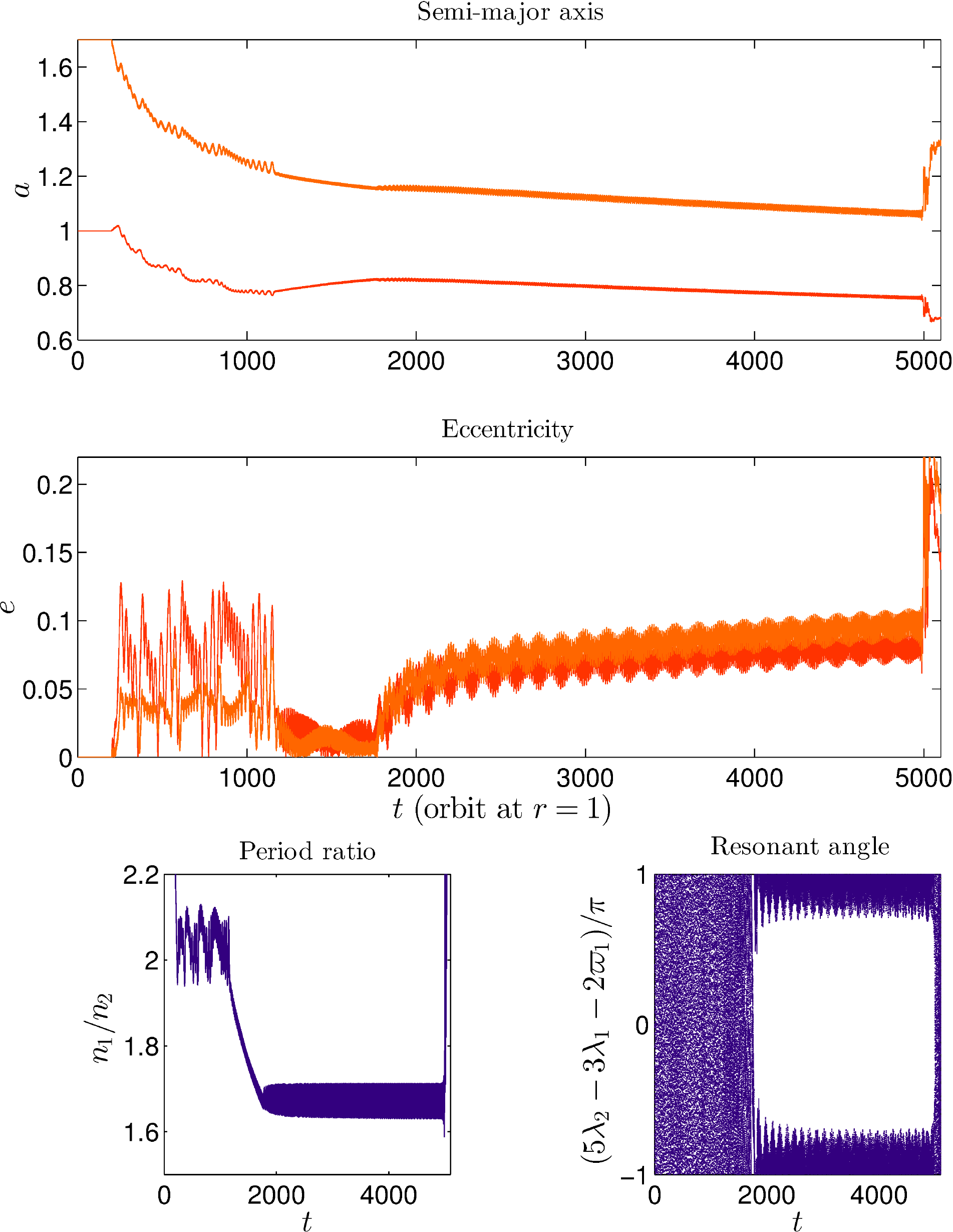}
\caption{Results for a 2D two planet run with the same parameters as for  the standard run,
except that the initial  disc model was modified such that the surface density
 was increased everywhere  by a factor of 5.
 The panels correspond to those of Fig. \ref{fig8}, except for the
 lowermost right hand panel which shows
 the resonant angle $5\lambda_2 - 3\lambda_1 - 2\varpi_1,$ appropriate for the 5:3 resonance. 
 Note that the middle panel shows the evolution of eccentricities, with that of the inner planet initially mostly smaller 
  but with the  curves for the two planets later overlapping. 
This run  eventually goes unstable after approximately
$5000$  inner planet orbits.}\label{fig17}
\end{figure}

\begin{figure}
\centering
\includegraphics[width=7.5cm]{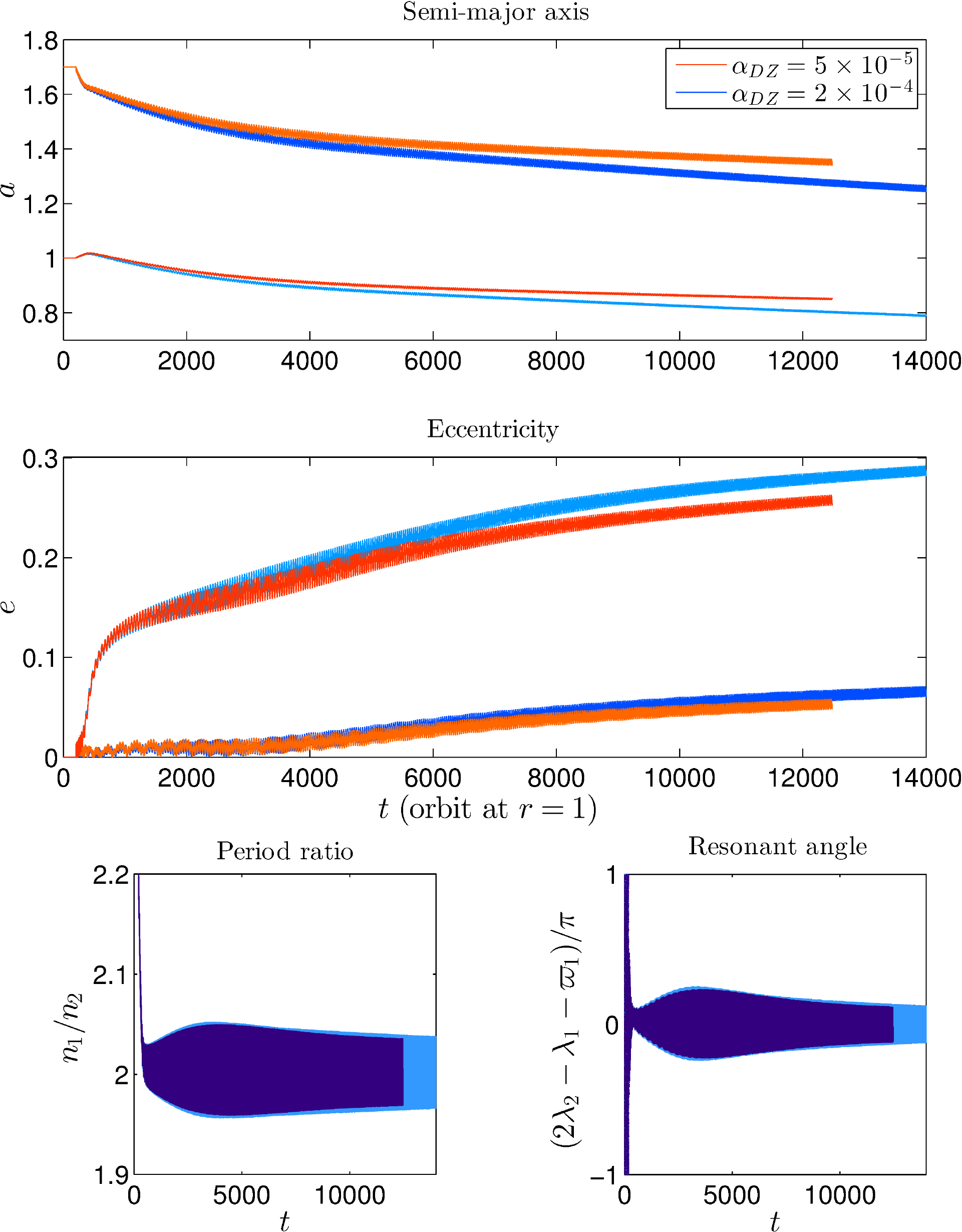}
\caption{A comparison  of  two 2D runs which were identical to the standard case
except that the values of $\alpha$ in the dead zone  were taken to be $2\times 10^{-4}$ and $5\times 10^{-5}$  respectively.
The panels correspond to those in Fig. \ref{fig8}.
The period ratio and resonant angle plots  overlap for these cases.
}\label{fig500}
\includegraphics[width=7.5cm]{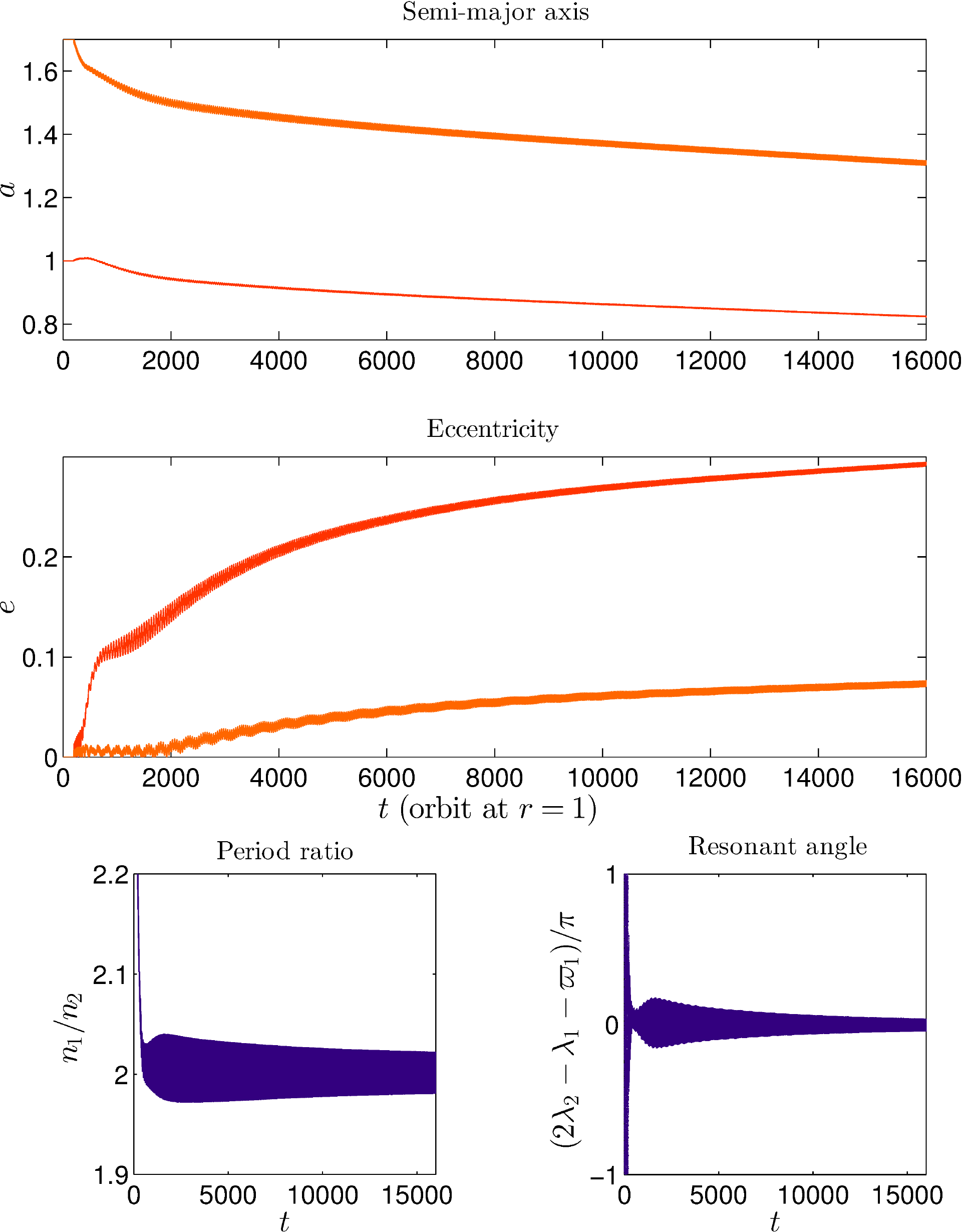}
\caption{Results for a  2D run which was identical to the standard case
except that the value of $\alpha$ in the active  zone  was taken to be  $10^{-2}.$  
The panels correspond to those in Fig. \ref{fig8}.}\label{fig5000}
\end{figure}

\begin{figure}
\centering
\includegraphics[width=7.5cm]{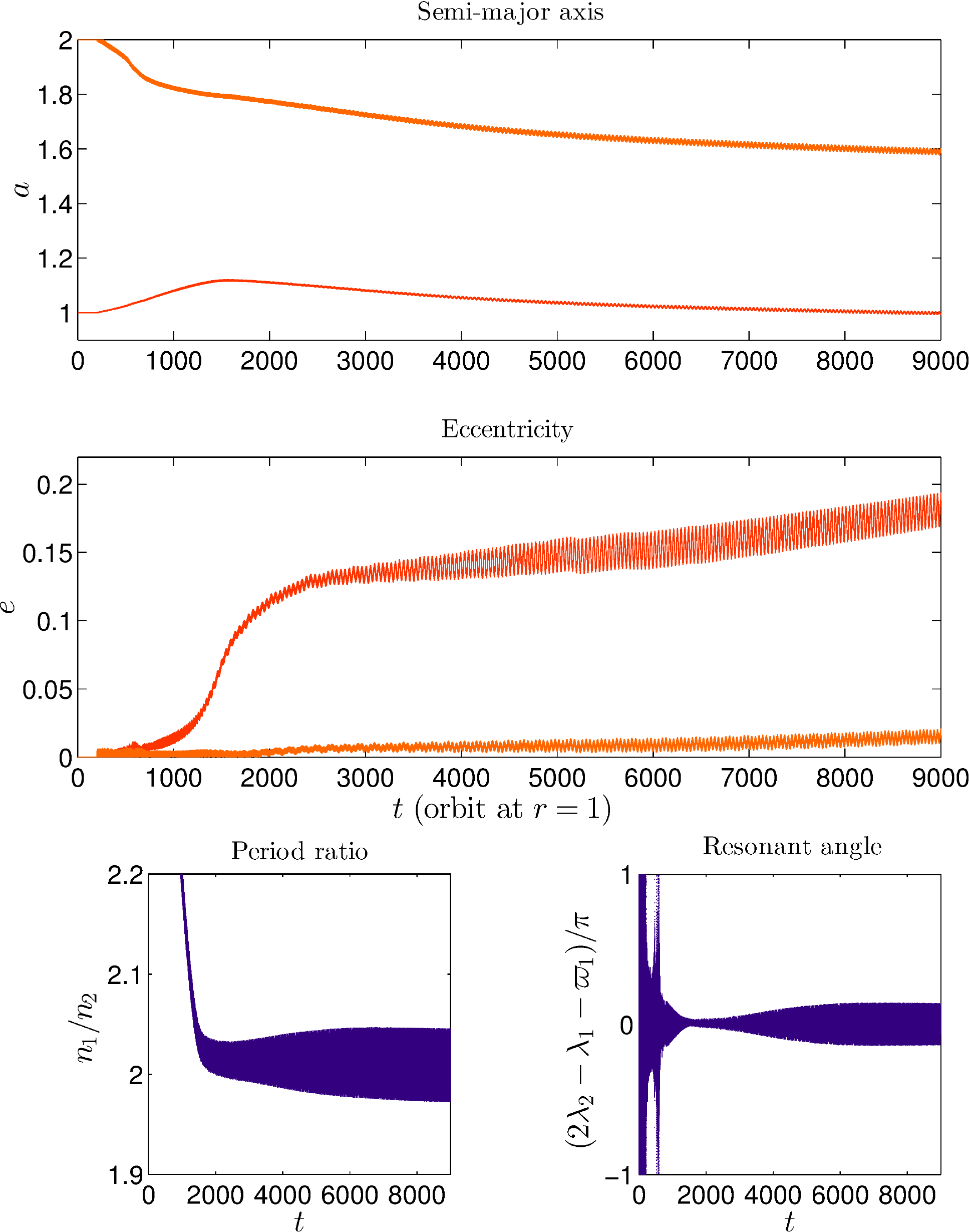}
\caption{As for the standard case shown in  Fig.\ref{fig8} except that  the surface density 
in the initial disc model is reduced through multiplication by a factor $0.78.$ 
 In addition,  the mass
of the outer planet which starts further out  at $r=2$ is increased by a factor of 2.}\label{fig20}
\centering \includegraphics[width=7.5cm]{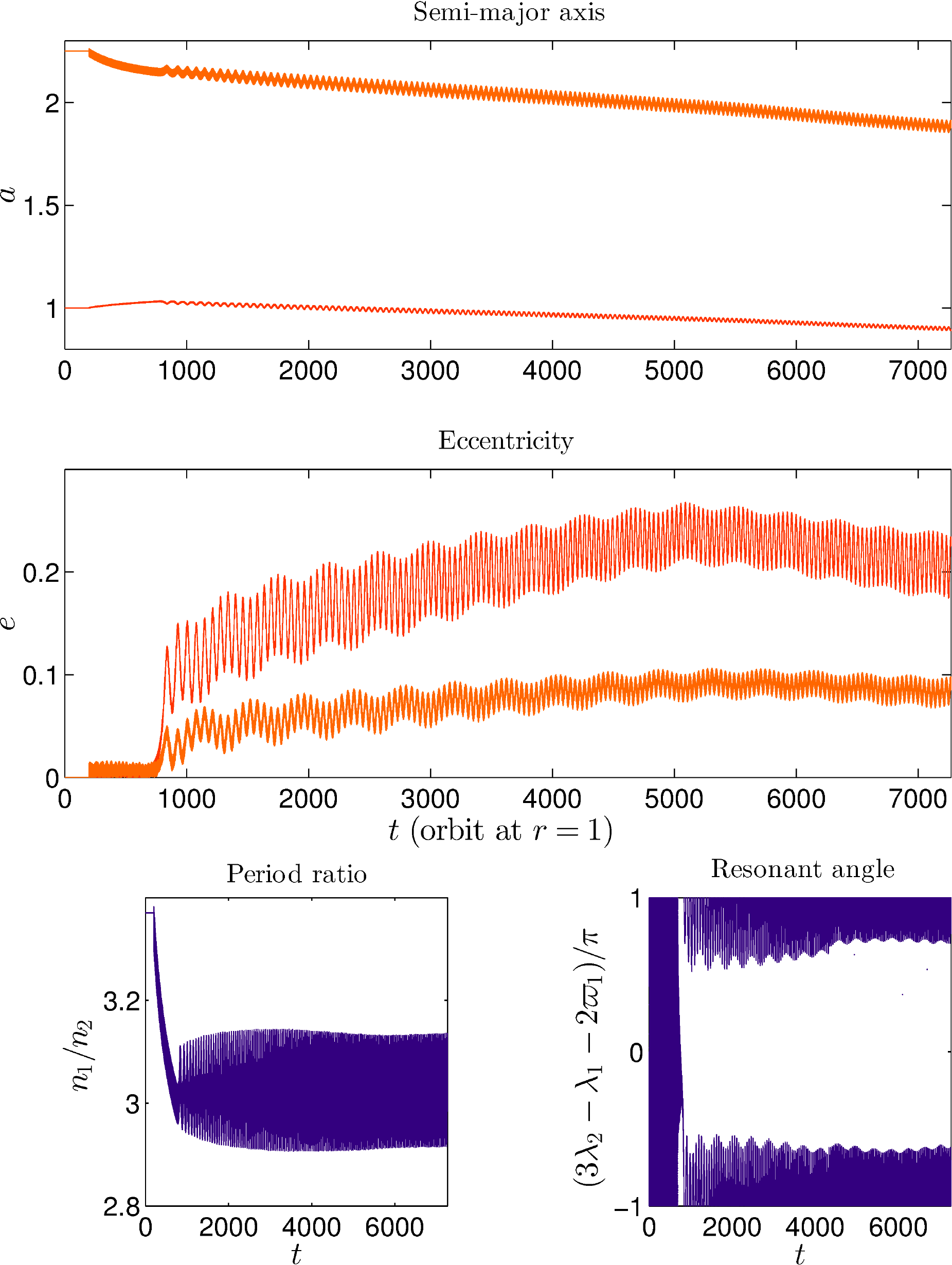}
\caption{Results for a 2D two planet run where the outcome is migration in a  3:1 resonance.
The parameters were chosen to correspond to the HD 60532 system (see text), with the outer planet starting at $r=2.25$.
The panels correspond to those in Fig. \ref{fig8} except that the resonant angle plotted is
$3\lambda_2-\lambda_1 -2 \varpi_1.$} \label{fig170}
\end{figure}

\subsection{Increased disc surface density and the formation of a 5:3 resonance}
\label{IncSig}
{ In order to investigate the effect of increasing the disc surface density we performed
2D  two planet runs  with the same parameters as the standard run,
except that the disc model was modified such that the surface density
 was increased by a factor  of five.
From the discussion of the disc models in Section \ref{discmod}, this
corresponds to increasing the steady state accretion rate by the same factor. 

 The results for this case are illustrated in Fig. \ref{fig17}.
%and those for the latter  in Fig. \ref{fig13}. For each of these Figures
The  uppermost panel shows the semi-major axes, the middle panel the eccentricities, the left  lowermost panel shows  the period ratio and the right lowermost panel shows
 a resonant  angle.

  The  outer planet
initially  migrates rapidly in a type III migration regime while the  semi-major axis 
of the inner planet hardly  changes indicating an approximate balance between inward and outward migration 
torques. As a consequence the planets rapidly enter a 2:1 resonance which quickly  develops an  instability,
as indicated by the strong fluctuations in eccentricity and period ratio. The instability causes the planets to leave the resonance
at around  $t= 1000$ and  resume convergent migration.  During this phase the inner planet migrates slowly outward
on account of the action of the inner disc and a deeper common gap.
  The planets enter a 5:3 resonance at $t \sim 2000.$  
This persists  until time $t \sim 5000$.} Accordingly
the resonant angle shown is
$5\lambda_2-3\lambda_1-2\varpi_1.$ This librates  about $\pi$ once the resonance forms. 
The angle $5\lambda_2-3\lambda_1-2\varpi_2$ shows very similar behaviour.
Although the estimated inward migration time  $-r/{\dot r} \sim 50,000$ inner planet { initial} orbital periods
while the system is in  5:3 resonance  this run  eventually goes unstable after approximately
$5000$ orbits culminating in the two planets undergoing a scattering (see also Lee et al. 2009  
for an approach based on N body methods). Thus observing a system in 5:3 resonance is unlikely.
{ We remark that although  the planet mass ratios differed and the transition was from a 2:1 resonance to a 3:2 resonance,  
qualitatively similar behaviour  to that described above,  up until the instability at late stages,  was  found  in  simulations of Rein et al. (2010) (see their Fig. 6).}

{ In order to investigate further the effect of increasing  disc mass we  performed an additional simulation for which the disc mass was increased by a factor of ten.
In this case the surface density  is large enough that the Toomre criterion for gravitational instability is close to being marginal at the outer boundary, that being potentially
the most unstable location.
As self-gravity has been neglected, this  disc model provides the  maximum  strength of the disc planet interaction that we can consider.
We investigate it in order to check that it is the model for which  the closest commensurability
is attained as indicated by the discussion at the beginning of Section \ref{stdrun}.

This case behaves similarly to the previous one except that the initial migration phase is more rapid and such that the system passes
straight through the 2:1 resonance before going into an unstable 5:3 resonance. After  $2000$ orbits it then undergoes a transition to a stable
3:2 resonance where it remains until 10,000 orbits after  the  start of simulation. 
While  the system in the 3:2 resonance the  mean migration rate is approximately $50\%$ faster than for the previous case. 
As expected  the closest commensurability was  attained in this case.}

\subsection{The effect of changing the magnitude of the viscosity}
In order to examine the effect of  changing the magnitude of the viscosity, 
 a  comparison of the semi-major axis evolution for different 2D runs is given in Fig. \ref{fig500}.
For these runs the value of $\alpha$ in the outer inactive region  was increased and decreased by a factor of two as compared to the
standard run, while the surface density was scaled such as to maintain the same steady state accretion rate as for the standard run.  The inner active disc  region was  the same as in the  standard case.
The quantities plotted in the different panels are as in Fig. \ref{fig8}.
We see that  the migration rate decreases with decreasing viscosity as expected but  somewhat  less steeply
than linearly. This is  because  the migration rate  has a dependence  on the disc  surface density profile
which  takes a long time to evolve,  especially in the case with the smallest viscosity.

In Fig. \ref{fig5000} we present a 2D run for which the initial conditions were as in the standard case
except that the value of $\alpha$ in the inner active region was increased from $10^{-3}$ to $10^{-2}$.
The quantities plotted in the different panels are as in Fig. \ref{fig8}. The results on this case
are very similar to those obtained for the standard run. We remark that the planets
are migrating in the inactive part of the disc.  Accordingly,  this result indicates  that the inner active disc
plays only a minor role in these circumstances.

\subsection{A case with a more massive outer planet}
  We have also considered the effect of increasing the mass of the outer planet.
In Fig. \ref{fig20} we give the results for a simulation for which the mass ratio
 of the outer planet was increased by a factor of two and its starting distance was increased from 
$1.7$ to $2$ in dimensionless units otherwise initial conditions are as in the standard model.
 However, the surface density was everywhere taken to be
 seventy eight percent of the standard value. This was done so as to make the disc mass in the
neighbourhood of the outer planet the same as in the standard case. In addition, the grid outer boundary was shifted from 3.75 to 4.4 such that it was kept distant enough from the outermost planet. Bearing in mind the observational uncertainties, 
the mass ratio being larger
 for the outer planet in this system  allows it to  potentially resemble  the HD 6805 system.
This run attains a 2:1 resonance and migrates inward with the two resonant angles  ultimately librating around zero
and such that $-r/{\dot r}\sim 10^5$ inner planet initial orbital periods. 
{ This characteristic time is  close  to that found for the
  standard run and so a very similar discussion will follow (see Section \ref{Disc}).}

\subsection{The formation of a 3:1 resonance }
We have also studied a case with planet masses chosen to correspond to the HD 60532 system 
(see table \ref{table1}) where the planets have been found to be in a  3:1 resonance
 (see Laskar  \& Correia, 2009).

The results of the simulations are shown in Fig. \ref{fig170}.
The quantities plotted in the different panels correspond to those in Fig. \ref{fig8}
except that the right lowermost  panel shows the   resonant angle
$3\lambda_2-\lambda_1-2\varpi_1.$
This is found to exhibit large amplitude librations
about $\pi$ as does the angle $3\lambda_2-\lambda_1-2\varpi_2.$
The behaviour of the angles that we find here is similar to that 
presented  by Laskar \& Correia (2009).
{ Note that the  two planets speed up their joint inward migration slightly  between $t=5000$ and $t=7000$ 
while their  mean eccentricities decrease slightly.  As  the rate of growth of eccentricity increases with the rate of convergent migration
as the planets become closer to resonance (e.g. Papaloizou 2003, Baruteau \& Papaloizou 2013), this is an indication that the planets are tending to converge more slowly
rather than there being an increase in the damping rate of their eccentricities.  That  in turn can be traced to
 a tendency of the inner planet to increase its rate of migration inwards  at  late times unlike in  most  other 2D cases where an inner planet is pushed
 by an outer planet.
 Note that in this case the difference is that the inner planet is more massive and so more effective  at clearing away the  inner disc which opposes
 its inward migration, 
 %( see eg.Fig. \ref{figsingle})
 }

We remark that this system has been considered by S\'andor  \& Kley (2010).
They considered a disc with the same aspect ratio 
and approximately the same mass interior to the inner planet as considered here. However the viscosity
 parameter $\alpha = 0.01$ throughout  corresponding  to a much larger viscosity than we consider for our standard model disc.
They found that the  planets attained a 3:1 resonant configuration with $-r/{\dot r}\sim 4500$ orbits.
In our case this time is extended to 60,000 initial inner planet orbits. { This is discussed further below.}

\section{Summary and Discussion} \label{Disc}
\label{sec:disc}

In this paper we have performed 2D and 3D simulations of pairs of giant planets
that have attained a mean motion resonance  in a protoplanetary disc.
We considered disc models both with an inner active region and an outer inactive region
with lower effective viscosity as well as disc models incorporating only one of these regions.
Different magnitudes for the viscosity in both regions were considered. This was found 
to have only minor effects on the results as long as the surface density was
scaled such as to maintain the same steady state accretion rate -- or equivalently, 
outward directed angular momentum flux, and this scaling did not result in  the surface density becoming
so small that the planet mass dominated its local neighbourhood. 
Disc models with a range of masses corresponding to a range of accretion rates were considered,
the smallest corresponding to the late stages of the protoplanetary disc life time.
Simulations were run for up to 20,000 initial orbits of the inner planet.

\subsubsection*{Maintenance of a 2:1 commensurability}
 When  the mass ratio for both planets  was  $10^{-3}$, a  2:1 commensurability was maintained 
 for small enough disc masses.  For our  standard case with a  low mass disc  and  a  corresponding  steady state  accretion rate 
 of $6\times 10^{-10}\,M_{\odot}\, \mbox{yr}^{-1},$  
  the  inward  migration time   $-r/{\dot r} \sim 1.5\times 10^5$ inner planet initial orbital periods is  a characteristic viscous time scale and so corresponds to standard type II migration.
   
  Noting that observed parameters are somewhat uncertain, { those for} this model  may  approximately correspond  to 
 those for HD 155358 and  24 Sextantis (see  table \ref{table1}) if the inner planet semi-major axis
 is respectively  taken  to be 0.64 au and 1.33 au { respectively.}
  The inward   migration  times  are  then  respectively $\sim 0.77\times 10^5\, \mbox{yr}$ and  $ 2.3\times 10^5\, \mbox{yr}$  
  which are relatively short compared to a characteristic  protoplanetary  disc life time.
{ For illustrative purposes, let us  assume}  the planets have migrated   in resonance for a time, $t,$ and  use
 the  scaling procedure given in section \ref{scaling}
 to  estimate the initial radius  of the inner planet. { For} either of the two examples {  we  obtain} 
$r = (t/(1.5\times 10^5\, \mbox{yr}))^{2/3}\, \mbox{au} \sim 3.5\, \mbox{au}$ for $t= 10^6\, \mbox{yr}.$ 
We note that this is beyond the ice line, with location estimated at about 2.7 au from a Solar mass star (see e.g. Martin \& Livio 2013). But we emphasise that the scaling used restricts the disc model such that
$\Sigma \propto r^{-2}.$ While this might be relaxed to some extent  the surface density  cannot 
exceed this projection by a large amount because a commensurability with period ratio
closer to unity would be formed (see below).

However, we note  that the quoted eccentricity for the inner planet in  HD 155358 is $0.17\pm 0.03$ (Robertson et al. 2012b).
The mean value is exceeded for the standard run at $4000$ orbits after going into  resonance with the implication that 
the planets could not have been in resonance longer than this time.  
If this is the case  the  above discussion would have to be 
modified  to allow  the planets  to migrate independently from larger radii before converging on to resonance close to their final
locations. This is likely to  need to be considered for different possible exterior disc models and  in addition the planets
may have built up their masses as they went (see e.g. Tadeu dos Santos et al. 2015). These considerations are beyond the scope of this paper.
Nonetheless,  because the migration rates for  single planets and  the resonantly coupled planets are in general  similar, the estimated starting radii
would also be similar for disc models that are similar to those we considered.
But note that the  attained eccentricities depend on the eccentricity damping rates which depend on the 
details of the disc model (see Crida et al. 2008).  For example,  we found that  for the same amount of relative resonant  migration, the entirely inactive disc model
led to smaller eccentricities while the 3D layered model led to larger eccentricities.
Thus it is important to note that there is  uncertainty as to how long the planets could have been in resonance. 
In the same context we comment that  migration in the completely active disc model was slower by a factor $\sim 1.6$ compared to the standard
case on account of its lower mass, that being  determined so as to maintain the same steady state accretion rate as in the standard case.
{ Furthermore the potential importance   a residual  inner  gaseous disc for  damping the eccentricity  of  the inner   planet and so  preventing the eccentricities
of both planets  from continuing to increase  in the later stages of the orbital evolution 
has been stressed by Crida et al. (2008).  In addition  Murray et al. (2002)  indicate   that a  residual  disc of planetesimals could produce  a similar  effect.}
 
 We  undertook 3D simulations that incorporated  consideration of the vertical structure of the disc.
 Both models that adopted a viscosity that was independent of  $\theta$ and layered models for which 
 a viscosity was only applied in the upper portion of the $\theta$ domain were considered.
The orbital evolution that was obtained was  found to be in  good agreement with  that obtained from  corresponding 2D simulations.
One effect seen in the 3D simulations that cannot be recovered from the 2D simulations is the vertical flow towards the
mid-plane in the interior neighbourhood of  the planet. However, because this occurs in the gap region where the density is very low,
this does not lead to significant departures from the 2D results for the orbital evolution.

 In order to consider a system resembling  the HD 6805 system, we performed a simulation identical to the standard 
 one except that the mass of the outer planet was increased by a factor of two.
This behaved like the standard case with maintenance of a  2:1  commensurability and an
inward migration rate   $-r/{\dot r}\sim 10^5$ inner planet initial orbital periods. 
Using the same scaling argument as above,   the inner planet can be estimated to  start at a radius 
being $\sim 5.5\, \mbox{au}$   if  resonant migration is assumed  for $t=10^6\, \mbox{yr}.$

\subsubsection*{Increasing planet mass and the formation of a 3:1 resonance }
We have also studied a case with planet masses chosen to correspond to the HD 60532 system 
which  has the larger  planet mass ratios  $2.2 \times 10^{-3}$ and $5.2 \times 10^{-3}$
(see table \ref{table1}).   These  planets are observed  to be in a  3:1 resonance
 (Laskar  \& Correia, 2009).
In our simulation,  the  planets attained a  3:1 resonant configuration with $-r/{\dot r}\sim 6\times 10^4$ initial inner orbits.
 { For the purposes of an illustrative discussion, if we assume that  the scaling to larger radii discussed in Section
 \ref{scaling} applies,
 we find that if the system arrived in its present location after having undergone
inward migration in resonance for $10^6\, \mbox{yr}$, it should have started with the inner planet at 
 an orbital radius of $\sim 7.4$ au.
For a shorter  evolution time of $4\times 10^5\, \mbox{yr}$ the corresponding starting location  shifts to 4 au.
However, note that as the orbital configuration  obtained in the simulation  is like  that observed,
 the two planets may have only spent a relatively small time in resonance,  
 comparable to  our simulation  run  time  of $\sim 6000$ orbits.
In that case the planets could have migrated independently, starting  at  initial
 radii that did not differ by a large factor  on account of the single planet migration rates being 
 comparable. If this is the case, detection of the system in resonance would be  unlikely. On the other hand only 
one system of this kind is currently  known.}

\subsubsection*{Effect of increasing disc mass}
When the surface density or  equivalently the steady state accretion rate was increased, the character of the migration of the resonantly coupled
pairs of Jupiter mass planets changed. When it was increased by a factor of 5 the planets are found to enter a  5:3
resonance which became unstable after about 5000 orbits 
leading to  a planet-planet scattering  as was found by Lee et al. (2009)   who adopted an N body approach.
%When the surface density was increased by a factor of $25$ the planets attained a  3:2 resonance
%which was stable but with a significantly reduced  inward  migration  time  $\sim 1.5\times 10^4\, \mbox{yr}.$ 
%Applying the scaling argument used above would suggest a starting radius $\sim 17\, \mbox{au}$
%for migration times $\sim 10^6\, \mbox{yr}.$  This coupled with
 The unstable character of the 5:3 resonance 
makes the observed occurrence of such a  resonances less  likely than a 2:1 resonance. 

If  pairs of  planets formed at a few au and then migrated to their present locations
with the inner planet being at around 1 au (such as for the HD 155358 and  24 Sextantis systems) while maintaining
a 2:1 commensurability  for a characteristic time comparable to the disc life time,
the disc should have a  low mass as might occur  during the later stages of a protoplanatary disc
lifetime. {  We have found that  a disc with significantly larger mass produce an unstable 5:3 resonance 
resulting in  its  observed occurrence being less likely.  Although 3:2 resonances may be produced in other situations
(eg. Rein et al. 2010 and see the end of Section \ref{IncSig} above) characteristic evolution times are again short.}
 
%We mentioned in Section \ref{IncSig} that when the surface density was increased by a factor of $25$ the planets 
%had a brief initial period during which they migrated outwards together.
%This aspect may be related to the mechanism outlined by Masset \& Snellgrove (2001).
{ Finally  consideration of   systems containing a pair of giant planets with the innermost one being significantly more massive, for which 
the mechanism outlined by Masset \&  Snelgrove (2001)  may operate more efficiently  should be undertaken but is beyond the scope of this paper.
 A recently discovered system of this kind is HD 204313 (Robertson et al. 2012a). This contains an inner planet of mass $3.55\, M_J$ with semi-major axis $3.04\, \mbox{au}$ and an outer planet of mass $1.68\, M_J$ with semi-major axis $3.93\, \mbox{au}$, the pair being in or close to a 3:2 commensurability.
Accordingly, this will be the focus of a future study.}

\section*{acknowledgment}
{ We thank the anonymous referee for useful comments and suggestions which greatly enhanced the quality of this paper. QA was supported by ENS Cachan and thanks warmly DAMTP, University of Cambridge for their hospitality.}

%\newpage

% Don't change these lines
\bsp	% typesetting comment
\label{lastpage}
\end{document}